\documentclass[letterpaper,11pt]{article}
\pdfoutput=1

\usepackage{xspace}
\usepackage{slashed}
\usepackage{array,multirow}
\usepackage{graphicx}
\usepackage{graphics}
\usepackage{epsfig}
\usepackage{amsfonts}
\usepackage{amsmath}
\usepackage{amssymb}
\usepackage{dsfont}
\usepackage{color}
\usepackage{xcolor}
\usepackage{cite}
\usepackage{pdflscape}
\usepackage{pifont}
\usepackage[utf8]{inputenc}
\usepackage{tikz} 
\usepackage{pgfplotstable}

\newcommand{\al}{\alpha}
\newcommand{\bt}{\beta}
\newcommand{\g}{\gamma}

\newcommand{\dt}{\delta}

\newcommand{\simu}{\sigma^{\mu\nu}}

\newcommand{\vL}{\ensuremath{\mathcal{L}}}

\usepackage{bm}  %

\newcommand{\beq}{\begin{equation}}
\newcommand{\eeq}{\end{equation}}
\newcommand{\bea}{\begin{eqnarray}}
\newcommand{\eea}{\end{eqnarray}}
\newcommand{\ben}{\begin{eqnarray*}}
\newcommand{\een}{\end{eqnarray*}}

\newcommand{\boldtau}{\mbox{\boldmath $\tau$}}

\newcommand{\boldpi}{\mbox{\boldmath $\pi$}}

\renewcommand{\vec}[1]{{\mathbf #1}} 

\newcommand{\bma}{\begin{pmatrix}}
\newcommand{\ema}{\end{pmatrix}}

\def\lixo#1{}

\def\slashchar#1{\setbox0=\hbox{$#1$}           
  \dimen0=\wd0                                    
  \setbox1=\hbox{/} \dimen1=\wd1                  
  \ifdim\dimen0>\dimen1                           
    \rlap{\hbox to \dimen0{\hfil/\hfil}}            
    #1                                             
  \else                                          
    \rlap{\hbox to \dimen1{\hfil$#1$\hfil}}        
    /                                           
 \fi}                                           %

\newcommand{\tb}{\bar \theta}

\newcommand{\Or}{\mathcal O}

\newcommand{\vp}{\varphi}
\newcommand{\sq}{^{2}}

\newcommand{\dslash}[1]{#1 \llap{/\kern-0.5pt}}
\newcommand{\Dslash}[1]{#1 \llap{/\kern+1.5pt}}
\newcommand{\DDslash}[1]{#1 \llap{/\kern+2.3pt}}
\newcommand{\dslashh}[1]{#1 \llap{/\kern+1pt}}

\newcommand{\Ex}[1]{\cdot 10^{#1}}
\newcommand{\nn}{\nonumber}

\begin{document}

\begin{titlepage}

\begin{flushright}
SI-HEP-2018-29 \\
QFET-2018-18\\
ACFI-T18-14

\end{flushright}

\vspace{2.0cm}

\begin{center}
{\LARGE  \bf 

The phenomenology of electric dipole moments in models of scalar leptoquarks

}
\vspace{2cm}

{\large \bf  W. Dekens$^{a,b}$, J. de Vries$^{c,d}$, M. Jung$^{e}$, K.K. Vos$^{f}$} 
\vspace{0.5cm}

\vspace{0.25cm}

{\large 
$^a$ 
{\it Theoretical Division, Los Alamos National Laboratory,
Los Alamos, NM 87545, USA}}

\vspace{0.25cm}
{\large 
$^b$ 
{\it 
Department of Physics, University of California at San Diego, 9500 Gilman Drive,\\ La Jolla, CA 92093-0319, USA}}

\vspace{0.25cm}
{\large 
$^c$ 
{\it 
Amherst Center for Fundamental Interactions, Department of Physics, University of Massachusetts Amherst, Amherst, MA 01003, USA}}

\vspace{0.25cm}
{\large 
$^d$ 
{\it 
RIKEN BNL Research Center, Brookhaven National Laboratory, Upton, New York 11973-5000, USA
}}

\vspace{0.25cm}
{\large 
$^e$ 
{\it Excellence Cluster Universe,
Technische Universit\"at M\"unchen, Boltzmannstr. 2, D-85748 Garching, Germany }}

\vspace{0.25cm}
{\large 
$^f$ 
{\it 
{%
    Theoretische Physik 1, Universit\"at Siegen,%
    Walter-Flex-Stra\ss{}e 3, D-57068 Siegen, Germany%
}}}

\end{center}

\vspace{1cm}

\begin{abstract}
\vspace{0.1cm}
We study the phenomenology of electric dipole moments (EDMs) induced in  various scalar leptoquark models. We consider generic
leptoquark couplings to quarks and leptons and match to Standard Model effective field theory. After evolving the resulting operators
to low energies, we connect to EDM experiments by using up-to-date hadronic, nuclear, and atomic matrix elements. We show
that current experimental limits set strong constraints on the possible CP-violating phases in leptoquark models. Depending on the
quarks and leptons involved in the interaction, the existing searches for EDMs of leptons, nucleons, atoms, and molecules all play a
role in constraining the CP-violating couplings. We discuss the impact of hadronic and nuclear uncertainties as well as the
sensitivities that can be achieved with future EDM experiments. Finally, we study the impact of EDM constraints on a specific
leptoquark model that can explain the recent $B$-physics anomalies.

\end{abstract}

\vfill
\end{titlepage}

\section{Introduction}

Leptoquarks are hypothetical particles that have the distinct property that they can transform quarks into leptons and vice
versa. They appear in various extensions of the Standard Model (SM) and recently have become more popular as certain leptoquark
models can reduce the tension between SM predictions and experimental data in semileptonic $B$ meson decays (see \emph{e.g.}
Refs.~\cite{Fajfer:2012jt, Buttazzo:2017ixm} and references therein; for a selection of recent papers discussing scalar leptoquarks see Refs.~\cite{Crivellin:2017zlb, Marzocca:2018wcf,Becirevic:2018afm}). The phenomenology of leptoquarks has been recently reviewed in great
detail in Ref.~\cite{Dorsner:2016wpm}.

One phenomenological aspect of leptoquarks which has been studied in less detail so far is their impact on permanent electric dipole
moments (EDMs). Nonzero EDMs of systems with a non-degenerate ground state break parity (P) and time-reversal (T) symmetry and by the
CPT theorem also CP symmetry. A nonzero EDM measured in a current or next-generation experiment would be a clear sign for a new
source of CP violation, as the only confirmed source of CP violation, the phase in the CKM matrix, predicts EDMs that are orders
of magnitudes below current and expected limits \cite{Pospelov:2013sca,Seng:2014lea,Mannel:2012qk}. For leptons or paramagnetic
systems such a measurement would translate directly into a discovery of New Physics (NP), while in systems like nucleons or diamagnetic atoms the new
source could in principle be of SM origin: this is due to the fact that the SM allows for another source of CP violation, dubbed
\emph{strong} CP violation and parametrized in form of the so-called QCD $\bar \theta$ term. Current EDM limits constrain $\bar \theta
< 10^{-10}$.
However, if nonzero EDMs are measured in several systems, their pattern can disentangle a small but nonzero $\tb$ term from genuine
beyond-the-SM (BSM) CP violation \cite{Dekens:2014jka}. 

The search for EDMs  has grown into a rich field with ongoing experiments to measure EDMs of muons, neutrons, various atoms and
molecules, and exciting efforts to measure EDMs of protons and light nuclei in electromagnetic storage rings, see e.g.
Refs.~\cite{Engel:2013lsa, Yamanaka:2017mef,Chupp:2017rkp} for recent reviews. In addition, EDMs can be probed by analyzing
decay patterns. This has been done for tau leptons in processes such as $e^+e^-\rightarrow\tau^+ \tau^-$ \cite{Inami:2002ah} and
$e^+e^-\rightarrow e^+e^- \tau^+ \tau^-$ \cite{Abdallah:2003xd} and is planned for strange and charmed baryons at
LHCb~\cite{Botella:2016ksl}.

In this work we investigate EDMs of the above-mentioned systems in scalar leptoquark models, while vector leptoquarks are left to future
work. Previous studies of EDMs in the context of leptoquark models \cite{Barr:1987sp, Geng:1990gr, Barr:1992cm, He:1992dc,
Herczeg:2003ag, Arnold:2013cva, Fuyuto:2018scm,Ramsey-Musolf:2017tgh}, have mainly focused on subsets of leptoquark interactions. Here we collect, rederive,
and extend the existing results to constrain all flavor-diagonal CP-violating couplings, including those to the second- and
third-generation fermions. We work out in  detail how to connect leptoquark models to EDM phenomenology in a modern effective-field-theory (EFT) framework. We discuss the relevance of QCD and electroweak renormalization-group evolution and use up-to-date hadronic, nuclear, and
atomic input, including a discussion of the hadronic and nuclear uncertainties.

All together we find that EDM searches set rather strong constraints on possible CP phases in leptoquark models. 
Since many flavor and collider observables provide only very weak constraints on imaginary parts, EDMs can provide important complementary
information. Given the importance of leptoquark models in the context of explanations for the $B$ anomalies it is interesting to study
their impact on EDMs, and, in particular, whether such models predict measurable signals in existing or future experiments. The
results obtained here should allow to answer such questions in a straightforward manner. As an explicit example, we apply our analysis
to a particular model that explains the $B$ anomalies \cite{Becirevic:2018afm}, showing that (future) EDM measurements can provide
relevant additional constraints.
 
This paper is organized as follows. In Sect.~\ref{classes} we list the possible scalar leptoquark representations following
Ref.~\cite{Dorsner:2016wpm}. We identify the two representations that are most interesting from the point of view of EDM phenomenology. In
Sect.~\ref{Match} we match these leptoquark models to CP-odd effective operators in SM-EFT and evolve the operators
down to a low-energy scale. We make the connection to EDM measurements in Sect.~\ref{sec:4} and discuss the required hadronic, nuclear,
and atomic matrix elements. In Sect.~\ref{pheno} we set constraints on the CP-violating phases and discuss the complementarity of
different EDM searches by studying scenarios with several nonzero CP-odd operators. In this section we also apply the EFT framework to 
the leptoquark model in Ref.~\cite{Becirevic:2018afm} that was constructed to explain the anomalies seen in $B$-meson decays, and
investigate the impact of the EDM limits. We conclude in Sect.~\ref{conclusion}.

\section{Classes of scalar leptoquarks}\label{classes}
Scalar leptoquarks couple in various ways to SM fields depending on their gauge representations. We follow the notation of Ref.\
\cite{Dorsner:2016wpm} and classify the leptoquarks by their symmetry properties under the SM gauge group. Six possible scalar
representations exist, two of which come with a particularly rich EDM phenomenology. 
In general, the Lagrangian describing leptoquark interactions can be written as
\begin{equation}\label{eq::L0}
\mathcal L_{LQ} = \mathcal L_{\mathrm{kin}} + \mathcal L_Y + \mathcal L_S\,,
\end{equation}
where the terms on the right describe the leptoquark kinetic terms, the interactions with SM fermions, and the scalar sector, respectively. The
kinetic term is simply given in terms of the gauge-covariant derivative which depends on the particular representation of the
leptoquark,
\begin{equation}
 \mathcal L_{\mathrm{kin}} = (\mathcal D_\mu S)^\dagger (\mathcal D^\mu S)\,.
\end{equation}
The Lagrangian for the scalar sector can be divided into a universal and  a non-universal piece,
\begin{equation}\label{eq::LS}
\mathcal L_S = (\mu_S^2 + \lambda_{HS}\varphi^\dagger \varphi)S^\dagger S + \lambda_S (S^\dagger S)^2 + \mathcal L'_S\,,
\end{equation}
where $\mathcal L'_S$ depends on the leptoquark representation. The universal part of $\mathcal L_S$ does not distinguish between the
different components in a leptoquark multiplet. This implies that all components obtain the same mass when $\mathcal L'_S=0$. The
interactions with fermions, $\mathcal L_Y $, cannot be written in a universal way and will be discussed  in more detail below.

Both the kinetic terms and the universal part of $\vL_S$ are CP-even, so that any CP violation has to come from either $\vL_Y$ or
$\vL_S'$.
In this work we focus on the former as it gives rise to a rich EDM phenomenology and the couplings to fermions play a role
in several explanations of the $B$-physics anomalies.
In particular, we will focus on the 
leptoquark representations that allow for both left- and right-handed couplings to fermions, as these  give significant contributions to CP-violating observables and thereby give rise to the most interesting EDM
phenomenology. 
Leptoquark models without this requirement still contribute to EDMs; however, in that case, the generation of a CP-violating phase necessarily involves a flavor change, which has to be reversed to induce EDMs by an additional non-diagonal weak interaction, rendering these contributions
much smaller. The requirement of both left-and right-handed couplings can also be avoided by introducing multiple leptoquarks. Although such scenarios can be certainly of interest, a complete analysis is beyond the scope of the current work  and we do not discuss them any further here.

The Lagrangian in Eq.~\eqref{eq::L0} constitutes a very minimal extension of the SM and, in general, the leptoquark will be accompanied by additional NP degrees of freedom. However, for the scalar leptoquark models the minimal extension is consistent in the sense that
Eqs.~\eqref{eq::L0}-\eqref{eq::LS} provide a renormalizable framework. In contrast, in vector leptoquark models the generation of a mass term similar to the one in Eq.~\eqref{eq::LS} necessitates the introduction of additional new fields. This renders the setup with vector leptoquarks highly model-dependent. We therefore refrain from discussing this class of models in the following.

\subsection{$R_2$ and $S_1$}

We start by discussing the interactions of the two scalar leptoquark representations with both left- and right-handed couplings. 
The first scalar leptoquark of this class falls into the $(3,2,7/6 )$ representation of $SU(3)_c\times SU(2)\times U(1)_Y$. The most
general form of the  interactions with the SM fermions can be written as
\bea\label{R2general}
\vL_Y^{(R_2)}=R_2^I\left( \bar u_R x_{RL} \epsilon_{IJ} L^J+ \bar Q^I x_{LR}^\dagger e_R\right) +{\rm h.c.}\,,\label{eq:R2lag}
\eea
where $I,J$ are $SU(2)$ indices and $x_{RL,LR}$ are  $3\times 3$ matrices in flavor space. 

The only other scalar representation that allows for  left- and right-handed couplings to fermions is $S_1\in (\bar 3,1,1/3 )$. The
allowed interactions are given by

\bea
\vL_Y^{(S_1)}=S_1^\g\left[\bar Q^{c,I}_\g y_{LL} \epsilon_{IJ} L^J + \bar u_{R\, \g}^cy_{RR}e_R
-\epsilon^{\al\bt\g}\bar Q^I_\al z_{LL}^\dagger \epsilon_{IJ} Q_\bt^{c,J}+\epsilon^{\al\bt\g}\bar d_{R\, \al}z_{RR}^\dagger
u_{R\, \bt}^c\right]+{\rm h.c.}\,,\label{eq:S1Lag}
\eea
where 
$\al,\,\bt,\, \g$ are $SU(3)_c$ indices,  $y_{LL,RR}$ and $z_{RR}$ are generic $3\times 3$ matrices in flavor space, while $z_{LL}$ is
a symmetric $3\times 3$  matrix. 
In principle, the interactions in Eqs.\ \eqref{eq:R2lag} and \eqref{eq:S1Lag} are defined in the
weak basis and have be rotated once we move to the mass basis after electroweak symmetry breaking (EWSB). In order to simplify this process, we choose a basis in which the
up-type quark and charged-lepton Yukawa matrices are already diagonal. Explicitly, we take the SM Yukawa couplings to be
\bea
-\vL_Y &=& \bar Q\vp Y_d d_R+\bar Q\tilde \vp Y_u u_R+\bar L\vp Y_e e_R+{\rm h.c.}\\
Y_u &=& \frac{\sqrt{2}}{v}{\rm diag}(m_u,\,m_c,\, m_t), \quad Y_e = \frac{\sqrt{2}}{v} {\rm diag}(m_e,\,m_\mu,\, m_\tau),\quad Y_d = V\, \frac{\sqrt{2}}{v}{\rm diag}(m_d,\,m_s,\, m_b)\,,\nn
\eea
where $V$ is the CKM matrix,  $\vp$ the Higgs doublet and $v$ its vacuum expectation value. This choice of basis implies that the couplings involving down-type quarks and/or neutrinos  obtain additional factors of CKM and/or PMNS matrix elements when moving to the mass basis (see Sect.\ \ref{sec:EWSB} for details). Instead, the interactions involving only up-type quarks and charged leptons are unaffected.

In their most general form the interactions of $S_1$  lead to baryon-number-violating interactions, while $R_2$ does not. Since the
experimental limits on proton decay  stringently constrain baryon-number violation (see, \emph{e.g.}, Ref.~\cite{Dorsner:2012nq}), we
will  assume baryon number to be conserved. This assumption has no implications for the $R_2$ leptoquark. For $S_1$ this implies that
either  $y_{LL}=y_{RR}=0$ or  $z_{LL}=z_{RR}=0$, and we will consider the two cases separately.

Finally, strong EDM constraints can naively be avoided by specifying that the leptoquarks only couple via one type of interaction, as all EDMs will be proportional to the combinations $\mathrm{Im}\,x_{LR}\, x_{RL}$, $\mathrm{Im}\,y_{LL}\, y_{RR}$, or
$\mathrm{Im}\,z_{LL}\, z_{RR}$. Such an assumption, however, can only be valid at one specific scale, since the second coupling will be generated via renormalization-group evolution due to Higgs exchange. 
For example, setting $x^{ab}_{RL}(\mu_H)=0$ at a certain scale $\mu_H$ leads to nonzero values at a different scale $\mu_L$ via (a similar relation holds for the $y_{LL,RR}$ couplings)
\begin{equation}
x^{ab}_{RL}(\mu_L) \sim \frac{y_a y_b}{(4\pi)^2}\,\log{\left(\frac{\mu_L}{\mu_H}\right)}\,x^{ab}_{LR}(\mu_H)\,,
\end{equation}
where $y_a$ and $y_b$ are the Yukawa couplings of the fermions involved in the interaction.  As renormalization requires both
interactions (assuming one is nonzero), one would generally expect both terms to be present with independent phases.

\section{Matching and evolution to low energies}\label{Match}
To assess the effects of Eqs.\ \eqref{eq:R2lag} and \eqref{eq:S1Lag} in low-energy observables, we need to evolve the corresponding
coefficients to low energies. Here we assume the leptoquarks to have masses well above the electroweak scale, $m_{LQ}\gg
v$, such that their effects can be described within an EFT. In fact, assuming that they predominantly decay to leptons and quarks of the same generation, searches at the LHC currently set limits of $m_{LQ}\gtrsim 1 -1.5$ TeV on scalar leptoquarks \cite{Aad:2015caa,Sirunyan:2018btu,Sirunyan:2018ryt,Sirunyan:2018vhk}.
To derive the low-energy Lagrangian at $\mu\simeq
1$ GeV, we first integrate out the leptoquarks at the scale $\mu\simeq m_{LQ}$ and match to effective CP-violating dimension-six
operators that appear in the SM-EFT Lagrangian \cite{Buchmuller:1985jz,Grzadkowski:2010es}. The resulting operators are evolved   to
the electroweak scale, using renormalization-group equations (RGEs), where heavy SM fields such as weak gauge bosons, the Higgs field,
and the top quark are integrated out.  This induces a set of $SU(3)_c\times U(1)_{\mathrm{em}}$-invariant operators which we
subsequently evolve down to a scale of $\mu \simeq 1$ GeV. To connect this low-energy Lagrangian to EDM observables we have to
evaluate  matrix elements of these operators using nonperturbative methods at the atomic, nuclear, and QCD levels. 
 The latter step is deferred to the next section, while we start with the matching to SM-EFT.

\subsection{Matching to CP-violating dimension-six SM-EFT operators}\label{sec:SMEFT}

Tree-level leptoquark exchange leads to several operators that contain CP-odd pieces:
\bea
\vL_{\psi^4} &=& C_{lequ}^{(1)\, abcd}(\bar L_a^I  e_{R_b}) \epsilon_{IJ}(\bar Q_c^J  u_{R_d})+ C_{lequ}^{(3)\, abcd}(\bar L_a^I  \simu e_{R_b})\epsilon_{IJ}\,(\bar Q_c^J  \sigma_{\mu\nu} u_{R_d}) \nn\\
&&+C_{quqd}^{(1)\, abcd}(\bar Q_a^I  u_{R_b}) \epsilon_{IJ}(\bar Q_c^J  d_{R_d})+C_{quqd}^{(8)\, abcd}(\bar Q_a^I t^a u_{R_b}) \epsilon_{IJ}(\bar Q_c^Jt^a  d_{R_d})
+{\rm h.c.}\,,
\label{eq:EFT4fermion}
\eea
where $t^a$ are the $SU(3)_c$ generators and $a,b,c,d$ are generation indices. Additional operators are induced at loop level,
\bea
\vL_{\rm dipole}&=&
\sum_{f=u,d,e}\left(C_{fB}\mathcal O_{fB}+C_{fW}\mathcal O_{fW} + {\rm h.c.}\right)+\sum_{q=u,d}\left(C_{qG}\mathcal O_{qG} + {\rm
h.c.}\right)+C'_{\tilde G}\mathcal O_{\tilde G}\label{eq::SMEFTdipolesJMT}
\\
&=&\bigg\{-\frac{g}{\sqrt{2}}\left[\bar Q \simu \tau^I W^I_{\mu\nu}\Gamma^u_{W}\tilde \vp u_R +\bar Q \simu \tau^I
W_{\mu\nu}^I\Gamma^d_{W} \vp d_R+\bar L \simu \tau^I W^I_{\mu\nu}\Gamma^e_{W} \vp e_R \right] \nn\\
&&-\frac{g'}{\sqrt{2}}\left[\bar Q \simu B_{\mu\nu}\Gamma^u_{B}\tilde \vp u_R +\bar Q \simu B_{\mu\nu}\Gamma^d_{B} \vp d_R+\bar L \simu B_{\mu\nu}\Gamma^e_{B} \vp e_R \right] \nn\\
&&-\frac{g_s}{\sqrt{2}}\left[\bar Q \simu t^a G^a_{\mu\nu}\Gamma^u_{G}\tilde \vp u_R +\bar Q \simu t^a G^a_{\mu\nu}\Gamma^d_{G} \vp
d_R\right]+{\rm h.c.}\bigg\} \nn\\
&&+C_{\tilde G}\frac{g_s}{6}f_{abc}\epsilon^{\al\bt\mu\nu}G^a_{\al\bt}G_{\mu\rho}^bG^{c\,\rho}_{\nu}\,,
\label{eq:SMEFTdipoles}
\eea
where we introduced\footnote{Note that $\epsilon^{\al\bt\mu\nu}$ is the completely anti-symmetric tensor with
$\epsilon^{0123}=+1$ and that this convention is opposite to that in Ref.~\cite{Grzadkowski:2010es}.}
\beq
C_{fB} = -\frac{g'}{\sqrt{2}}\Gamma_B^f\,,\quad C_{fW} = -\frac{g}{\sqrt{2}}\Gamma_W^f\,,\quad C_{qG} =
-\frac{g_s}{\sqrt{2}}\Gamma_G^q\quad{\rm and}\quad C'_{\tilde G} = -\frac{1}{3}g_s C_{\tilde G}
\eeq
for later convenience and
where $B_{\mu\nu}$, $W^I_{\mu\nu}$, and $G^a_{\mu\nu}$ are the  field strength tensors of the $U(1)_Y$, $SU(2)$, and $SU(3)_c$ gauge
groups, $g'$, $g$, and $g_s$ are the corresponding gauge couplings, and $\tau^I$ are the Pauli matrices.
The $\Gamma_{W,B,G}^{u,d,e}$ couplings are  $3\times 3$ matrices in flavor space. The first two lines of Eq.\ \eqref{eq:SMEFTdipoles}
represent the electroweak dipole moments,  the third line contains the color dipole-moments, while  $C_{\tilde G}$ in the fourth line
is the Weinberg operator \cite{Weinberg:1989dx}. Here we will focus on the combinations of the electroweak dipole moments that give 
rise to the electromagnetic dipoles after electroweak symmetry breaking. To identify this combination and simplify later
expressions, we introduce the following combinations of couplings:
\begin{align}\label{eq:edms}
\frac{Q_uy_{u_i}}{\sqrt{2}} C_{u_i}^{(\g)} &= -{\rm Im}\,\left[\Gamma^u_{B}+\Gamma^u_W\right]_{ii}=\frac{d_{u_i}}{e
v}\,,\quad &\frac{Q_dy_{d_i}}{\sqrt{2}}C_{d_i}^{(\g)} &= -{\rm
Im}\,\left[V^\dagger\left(\Gamma^d_{B}-\Gamma^d_W\right)\right]_{ii}=\frac{d_{d_i}}{e v}\,,\nn\\
\frac{Q_ey_{l_i}}{\sqrt{2}} C_{l_i}^{(\g)} &= -{\rm Im}\,\left[\Gamma^e_{B}-\Gamma^e_W\right]_{ii}=\frac{d_{l_i}}{e
v}\,,&&\nn\\
\frac{y_{u_i}}{\sqrt{2}} C_{u_i}^{(g)} &={\rm Im}\,\left[ \Gamma^u_{G}\right]_{ii}=\frac{\tilde d_{u_i}}{v},\quad
&\frac{y_{d_i}}{\sqrt{2}} C_{d_i}^{(g)} &= {\rm Im}\,\left[V^\dagger \Gamma^d_{G}\right]_{ii}=\frac{\tilde d_{d_i}}{v}\,,
\end{align}
where $Q_u$, $Q_d$, and $Q_l$ denote the electric charges of the corresponding fermions, $y_f = \sqrt{2}m_f/v$ the Yukawa
couplings, and $d_f,\tilde d_q$ are the conventional (chromo-)EDMs, defined via dimension-five operators.

Both the $S_1$ and $R_2$ leptoquarks give rise to the semileptonic operators in Eq.\ \eqref{eq:EFT4fermion} at tree level
\cite{He:1992dc,Barr:1992cm,Herczeg:2003ag, Fuyuto:2018scm}. At the scale $\mu=m_{LQ}$ we obtain 
\bea\label{eq:MatchFQsemil} 
C^{(1)\, abcd }_{lequ}&=& \frac{1}{2 } X_{abcd}^*+\frac{1}{2} Y_{abcd}^*\,,\qquad 
C^{(3)\, abcd}_{lequ} = \frac{1}{8} X_{abcd}^*- \frac{1}{8}Y_{abcd}^*\,.
\eea
 Only the di-quark couplings of $S_1$ induce the four-quark operators in Eq.~\eqref{eq:EFT4fermion}:
\bea\label{eq:MatchFQ}
 C^{(1)\, abcd }_{quqd}&=&-\frac{N_c-1}{N_c} Z_{abcd}\,,\qquad C^{(8)\, abcd }_{quqd}=2 Z_{abcd}\,,
\eea
where $N_c=3$ is the number of colors. In the above expressions we have defined the convenient combination of parameters
\begin{equation}\label{defLQ}
X_{abcd}\equiv \frac{\left(x_{LR}\right)^{bc}\left(x_{RL}\right)^{da}}{m_{R_2}\sq}\,,\,\, Y_{abcd}\equiv \frac{\left(y_{LL}\right)^{ca}\left(y_{RR}^*\right)^{db}}{m_{S_1}\sq}\,,\,\,  Z_{abcd}\equiv \frac{\left( z_{LL}^*\right)_{ac}\left( z_{RR}\right)_{bd}}{m_{S_1}\sq}\,.
\end{equation}
This implies $C^{(1,8)\, abcd }_{quqd}=C^{(1,8)\, cbad }_{quqd}$ because $z_{LL}$ is
symmetric in flavor space. Although Eqs.~\eqref{eq:MatchFQ} and \eqref{eq:MatchFQsemil} present the matching for general flavor
indices, only certain combinations will be most relevant for EDMs, namely those involving a single lepton (or down-type quark) flavor
and a single up-type quark flavor. It is therefore useful to define the following combinations:
\bea
X_{ab}\equiv {\rm Im}\, \left[ X_{aabb}\right]\,,\quad Y_{ab}\equiv {\rm Im}\, \left[ Y_{aabb}\right]\,,\quad Z_{ab}\equiv {\rm Im}\, \left[\sum_c V_{cb}^* Z_{caab}\right]\,.
\eea

\begin{figure}
\begin{center}
\includegraphics[scale = .9]{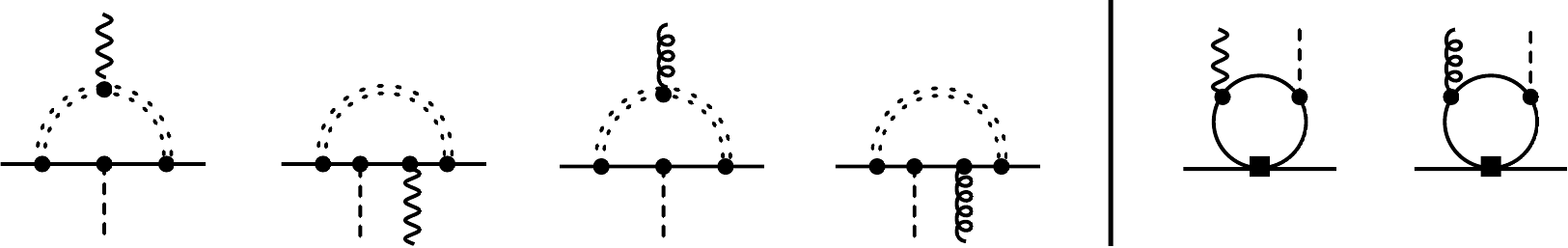}
\end{center}
\caption{Left panel: one-loop diagrams contributing to fermion (chromo-)EDMs  in the full theory including dynamical leptoquarks. Right
panel:  one-loop diagrams contributing to fermion (chromo-)EDMs in the EFT where leptoquarks have been integrated out. Solid lines
denote SM fermions, single dashed lines the Higgs field, double dashed lines leptoquarks, wavy lines photons, and curly lines gluons.
Circled vertices denote  vertices of dimension four, while squares denote effective vertices arising from higher-dimensional
operators. Only one topology for each diagram is shown.}
\label{oneloopEDM}
\end{figure}

\begin{figure}
\begin{center}
\includegraphics[scale = .8]{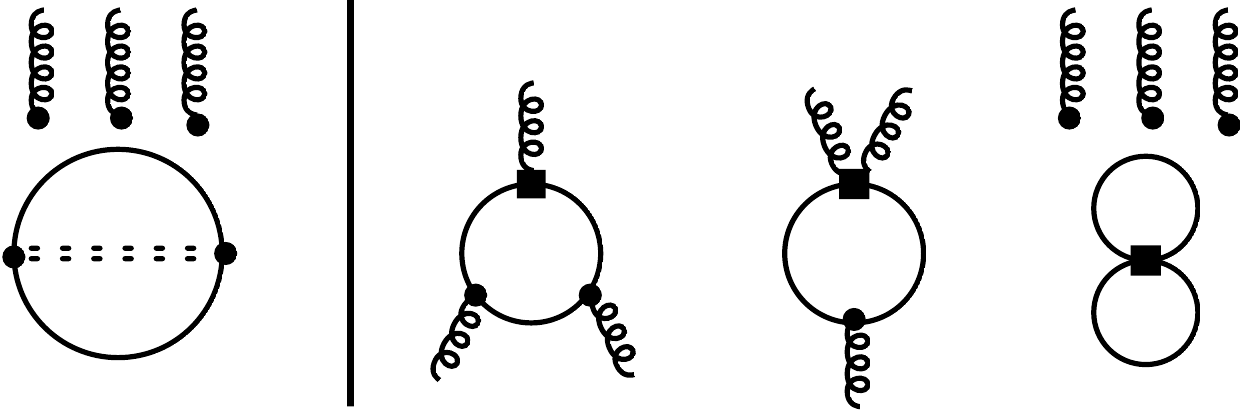}
\end{center}
\caption{Left panel: two-loop diagrams contributing to the Weinberg operator  in the full theory including dynamical leptoquarks. Right
panel:  one- and two-loop diagrams contributing to the Weinberg operator in the EFT where leptoquarks have been integrated out. The
disconnected gluons in the first and fourth diagram can attach to any internal line carrying color charge. Other notation as in
Fig.~\ref{oneloopEDM}.}
\label{twoloopEDM}
\end{figure}

The lepton-quark couplings of the two leptoquarks generate the (chromo-)EDMs  via the one-loop diagrams depicted in
Fig.~\ref{oneloopEDM}.
The (chromo-)EDMs are induced in the full theory via the first four diagrams where the external photon (gluon) couples to an internal
SM fermion or to the leptoquark. To match the full theory to SM-EFT we need to subtract the contribution from the fifth (sixth) diagram
that appears in the EFT through an insertion of $C_{lequ}^{(1,3)}$. These loops only contribute to EDMs, while the chromo-EDMs vanish
due to the color trace. After performing the matching calculation, we find at $\mu=m_{LQ}$ 
\bea\label{oneloopEDM1}
Q_l m_lC^{(\g)}_{l=e,\mu,\tau } &=& \frac{1}{(4\pi)\sq}\sum_{q=u,c,t}N_c m_q
\big[(Q_l/2+Q_q)X_{lq}+(Q_l/2-Q_q)Y_{lq}\big]+\dots\,,\nn\\
Q_q m_qC^{(\g)}_{q=u,c,t} &=&\frac{1}{(4\pi)\sq}\sum_{l=e,\mu,\tau} m_l\big[(Q_q/2+Q_l)X_{lq}+(Q_q/2-Q_l)Y_{lq}\big]+\dots\,,\nn\\
m_q C^{(g)}_{{q=u,c,t}} &=& -\frac{1}{2}\frac{1}{(4\pi)\sq}\sum_{l=e,\mu,\tau}m_l\big[X_{lq}+Y_{lq}\big]+\dots\,,
\eea
where the dots denote contributions suppressed by $(m_{q,l}/m_{LQ})^2$.  
In addition, there are two-loop diagrams in the full theory (left panel of Fig.~\ref{twoloopEDM}) that contribute to the Weinberg
operator, which were recently evaluated in Ref.\ \cite{Abe:2017sam}. We find that these contributions are canceled after
subtracting the one-loop diagrams in the EFT (the right panel of Fig.~\ref{twoloopEDM}):
\bea
C_{\tilde G}(m_{LQ}) = 0 + \dots\,,
\eea
where the dots denote higher-dimensional terms additionally suppressed by at least $(v/m_{LQ})^2$. Non-vanishing dimension-six
contributions to the Weinberg operator will appear at lower energies when heavy quarks are integrated out
\cite{BraatenPRL,Braaten:1990zt,Boyd:1990bx}. 

The above results differ from the literature in two ways. First of all, we consistently neglected contributions of dimension eight
and higher as these are additionally suppressed and beyond the scope of the present EFT approach. Second,
the loops in the full theory lead to contributions to the fermion EDMs proportional to $(m_{q,l}/M_{LQ}^2) \log (m_{q,l}\sq/M_{LQ}\sq)$
\cite{Barr:1987sp,Arnold:2013cva}. These logarithms do not appear in the matching once the diagrams in the EFT are subtracted, but will
be partially reintroduced when the effective operators are evolved to lower energy scales using the one-loop RGE. Importantly, this
approach allows to resum these logarithms.
In this way we incorporate the sizeable QCD corrections to the effective CP-violating operators that arise from evolution from $M_{LQ}$ to lower energies \cite{Dekens:2013zca}. 

Apart from the contributions of the lepton-quark couplings, the quark (color-)EDMs are also induced by one-loop diagrams involving the
di-quark couplings of $S_1$. These contributions again require a matching calculation, which leads to very similar expressions:
\bea\label{oneloopEDM2} 
Q_q m_q C^{(\g)}_{q=u,c,t} &=&\frac{1}{(4\pi)\sq} (N_c-1)\sum_{q'=d,s,b}m_{q'}(Q_{q'}-Q_q/2) Z_{qq'}+\dots\,\,,\nn\\
Q_q m_q C^{(\g)}_{q=d,s,b} &=&\frac{1}{(4\pi)\sq} (N_c-1)\sum_{q'=u,c,t}m_{q'}(Q_{q'}-Q_q/2)Z_{q'q}+\dots\,\,,\nn\\
m_q C^{(g)}_{q=u,c,t} &=& \frac{2}{(4\pi)\sq}\sum_{q'=d,s,b} m_{q'} Z_{qq'}+\dots\,,\nn\\
m_q C^{(g)}_{q=d,s,b} &=& \frac{2}{(4\pi)\sq}\sum_{q'=u,c,t} m_{q'} Z_{q'q}+\dots\,.
\eea
The di-quark couplings also induce two-loop contributions to the Weinberg operators but, as was the case for the quark-lepton
interactions, these contributions only appear at dimension eight or higher.

It should be mentioned that, apart from the interactions discussed above, additional operators are induced at loop level. Examples
are CP-odd Yukawa interactions  $Q_{fH}$ ($f=e,u,d$), as well as four-fermion operators such as $Q_{qu}^{(1,8)}$, in the notation of
\cite{Grzadkowski:2008mf}, which are induced through box diagrams involving two leptoquark exchanges. When considering the $X_{lq}$,
$Y_{lq}$, and $Z_{ud}$ couplings, these additional operators generate contributions to EDMs that are suppressed by additional loop
factors or small (SM) Yukawa couplings compared to the operators in Eqs.\ \eqref{eq:EFT4fermion} and \eqref{eq:SMEFTdipoles}. However,
in the case of flavor off-diagonal couplings the effects of operators like $Q_{qu}^{(1,8)}$ can become important; for example, its
$U_1U_2U_2U_1$ ($U_{1,2}\in\{u,c,t\}$, $U_1\neq U_2$) component is induced
by a combination of couplings $\sim X_{ll'U_1U_1} X_{ll' U_2U_2}^*$, whose effects are not captured by Eqs.\ \eqref{eq:EFT4fermion} and \eqref{eq:SMEFTdipoles} when $l\neq
l'$. Note, however, that in this case generally also sizable lepton-flavor-violating couplings would be induced which are tightly
constrained. A complete analysis of such cases would require extension of the above operator basis and is beyond the scope
of the current work.

\subsection{Renormalization group equations}\label{RGE}
Having derived the effective operators at the scale $\mu = m_{LQ}$, we now discuss the relevant RGEs needed to evolve these operators
to lower energies. 
 This process is complicated by the presence of the various hierarchies in the problem, \emph{i.e.} hierarchical masses and mixing
 angles, gauge couplings, and loop factors. In general all of the occurring operators undergo renormalization under
 QCD.\footnote{For some operators discussed in this work there can be a sizable additional contributions to the anomalous
 dimension matrix proportional to the top-quark Yukawa coupling. These have been recently discussed in Ref.~\cite{Buras:2018gto} and
 can be included analytically. However, we ignore these contributions in the following for simplicity.
}
While this changes the coefficients in some cases sizably, it does not generally change the hierarchies in the problem.
Contributions to the RGEs from weak interactions on the other hand can generate operators that enter the problem qualitatively
differently. The importance of these contributions is highly flavor dependent. For instance, leptoquark interactions between light
fermions only can contribute sizably already at tree-level, rendering terms generated via weak processes subleading. On the other hand,
EDMs induced by leptoquark exchange between just heavy particles typically receive their leading contributions from RGE mixing. The question
of the dominant low-energy contribution(s) therefore has to be evaluated on a case-by-case basis.
As the mixing pattern is rather complicated, in Fig.~\ref{Fig:RG} we show a flow diagram depicting the various matching and RGE contributions to low-energy CP-odd operators arising from the $R_2$ leptoquark interactions. 

We start with the self-renormalization of the four-fermion operators, which is governed by the following RGEs
\cite{Hisano3,An:2009zh,Dekens:2013zca}
\bea
\frac{d}{d\ln\mu} C_{lequ}^{(1)} &=& -6C_F\frac{\al_s}{4\pi}C_{lequ}^{(1)}\,,\qquad\quad \frac{d}{d\ln\mu} C_{lequ}^{(3)} =
2C_F\frac{\al_s}{4\pi}C_{lequ}^{(3)}\,,\nn\\
\frac{d}{d\ln\mu} \left(\vec C_{quqd}^{abcd}+\vec C_{quqd}^{cbad}\right)&=& \frac{\al_s}{4\pi} \bma -4C_F\frac{3N_c+4}{N_c} & 4
C_F\frac{2+N_c-N_c^2}{N_c\sq}\\
\frac{16+8N_c}{N_c} & 4 C_F-4\frac{2+N_c\sq }{N_c\sq}\ema \cdot \left(\vec C_{quqd}^{abcd}+\vec C_{quqd}^{cbad}\right)\,,
\label{eq:RGE4fermion}\eea
where $\vec C_{quqd}^{abcd} = (C_{quqd}^{(1)\, abcd},\, C_{quqd}^{(8)\, abcd})^T$ and $C_F = (N_c\sq-1)/(2N_c)$.
Neglecting the running due to the top Yukawa coupling, the RGEs for the semileptonic operators are independent of the flavor indices, while those for $C_{quqd}^{(1,8)}$ in principle still depend
on them \cite{Alonso:2013hga}. However, since $z_{LL}$ is symmetric, the
interactions of the $S_1$ leptoquark only contribute to the symmetric combination
for which the RGE is given in Eq.~\eqref{eq:RGE4fermion}.

At one loop, the four-fermion operators mix into the fermion dipole operators. For large scales $\Lambda$, these contributions are
expected to dominate the direct one-loop matching contributions in the last section, since they receive a logarithmic enhancement
$\sim \ln(\Lambda/\mu)$. While this is generally confirmed numerically in our analysis, the one-loop matching contributions are
important for two reasons: first of all, the logarithm is not very large when considering NP scales of one to a few TeV; consequently,
we find contributions of up to $50\%$ from the matching at a scale of 1~TeV. 
The second reason is specific to mediators carrying color charge, like the leptoquarks considered here: while the semileptonic operators do
not mix into chromo-EDM operators at the considered order, the latter are generated nevertheless at one loop in the
matching, see Fig.~\ref{oneloopEDM}. The chromo-EDM contributions are especially important in some cases, since they match onto the
Weinberg operator without an additional fermion mass factor. Hence this matching contribution can change the phenomenology
qualitatively, a fact which would be overlooked in an analysis only considering the logarithmic-enhanced RGE contributions.

The contributions from the semileptonic four-fermion operators to the lepton EDMs at one loop read
\bea
\frac{d}{d\ln\mu}  C_{l}^{(\g)} = -\frac{16N_c}{(4\pi)\sq}\sum_{q=u,c,t}\frac{m_q Q_q}{m_l Q_l}{\rm Im}\, C^{(3)\, llqq}_{lequ}\,.
\eea
These coefficients do not  run under QCD.

The (chromo-)EDMs of the quarks receive contributions from the four-fermion operators as well. In addition they, as well as the
Weinberg operator, evolve and mix under QCD.
The RGEs for the up-type dipole operators take the following form:
\bea
\frac{d}{d\ln\mu} \vec C_{q=u,c,t} &=& \frac{\al_s}{4\pi} \g \cdot \vec C_q+ \frac{1}{(4\pi)\sq}\sum_{l=e,\mu,\tau}\frac{m_l Q_l}{m_q Q_q}\g_{FF}{\rm Im}\, C^{(3)\, llqq}_{lequ}\nn\\
&&+\frac{1}{(4\pi)\sq}\sum_{q'=d,s,b}\frac{m_{q'}}{m_q }\, \g_{FQ}\cdot {\rm Im}\, \sum_{i=u,c,t}V_{iq'}^*\vec C_{quqd}^{iqqq'}\,,\label{eq:upRG}
\eea
where $\vec C_q = (C_q^{(\g)},\,C_q^{(g)},\, C_{\tilde G})^T$. Instead, for the down-type dipole moments we have
\bea
\frac{d}{d\ln\mu} \vec C_{q=d,s,b} &=& \frac{\al_s}{4\pi} \g \cdot \vec C_q+\frac{1}{(4\pi)\sq}\sum_{q'=u,c,t}\frac{m_{q'} }{m_q }\, \g_{FQ}\cdot {\rm Im}\, \sum_{i=u,c,t} V^*_{iq}\vec C_{quqd}^{iq'q'q}\,.
\label{eq:downRG}
\eea
Here the CKM elements appear due to our use of $C_{d_i}^{(\g,\, g)}$ and $m_{d_i}$, which are the (C)EDMs and masses of the quarks in
the mass basis, while $\vec C_{quqd}^{abcd}$ were defined in the flavor basis. The combinations in Eqs.\ \eqref{eq:upRG} and
\eqref{eq:downRG} are proportional to $Z_{qq'}$ after taking the matching in Eq.~\eqref{eq:MatchFQ} into account.

The QCD evolution in these equations is dictated by $\g$ which is given by
\cite{Degrassi:2005zd,Weinberg:1989dx,Wilczek:1976ry,BraatenPRL}
\bea
\g = \bma 8C_F &- 8 C_F & 0 
\\
0 &16 C_F -4N_c &2N_c 
\\
 0 & 0& N_c+2 n_f+\bt_0 
\ema \,,
\eea
where $n_f$ is the number of active flavors and $\bt_0 = (11 N_c-2n_f)/3$. Instead, the mixing of the four-fermion operators with the
dipole moments is determined by $\g_{FF,FQ}$ \cite{Hisano3,Dekens:2013zca, Cirigliano:2016njn,Jenkins:2013wua}:
\bea
\g_{FF} = -16\bma 1 \\0\\0\ema ,\qquad \g_{FQ} = \bma 2 \frac{Q_{q'}}{Q_q}& 2C_F  \frac{Q_{q'}}{Q_q}\\ -1 & N_c\\0&0\ema\,.
\eea

\subsection{Below the electroweak scale}\label{sec:EWSB}
Below the electroweak scale we integrate out the Higgs, $W^\pm$, and $Z$ bosons, as well as the top quark, and rotate to the mass basis. For the four-fermion operators in Eq.\ \eqref{eq:EFT4fermion}, the only effect is the removal of operators involving the top quark from the EFT and the appearance of several CKM and PMNS elements
\bea\label{eq:Broken4Fermion}
\vL_{\psi^4} &=& C_{lequ}^{(1)\, abcd}\left[U^*_{aa'}V^*_{cc'}\left (\bar \nu_{L_{a'}} e_{R_b}\right )\left (\bar d_{L_{c'}} u_{R_d}\right )-\left (\bar e_{L_a} e_{R_b}\right )\left (\bar u_{L_c} u_{R_d}\right )\right]\nn\\
&&+C_{lequ}^{(3)\, abcd}\left[U^*_{aa'}V^*_{cc'}\left (\bar \nu_{L_{a'}}\sigma_{\mu\nu} e_{R_b}\right )\left (\bar d_{L_{c'}} \sigma^{\mu\nu}u_{R_d}\right )-\left (\bar e_{L_a}\sigma_{\mu\nu} e_{R_b}\right )\left (\bar u_{L_c} \sigma^{\mu\nu}u_{R_d}\right )\right]\nn\\
&&+C_{quqd}^{(1)\, abcd}V^*_{aa'}\left[\left (\bar u_{L_{c}} u_{R_b}\right )\left (\bar d_{L_{a'}} d_{R_d}\right )-\left (\bar d_{L_{a'}} u_{R_b}\right )\left (\bar u_{L_c} d_{R_d}\right )\right]\nn\\
&&+C_{quqd}^{(8)\, abcd}V^*_{aa'}\left[\left (\bar u_{L_{c}} t^au_{R_b}\right )\left (\bar d_{L_{a'}}t^a d_{R_d}\right )-\left (\bar d_{L_{a'}} t^au_{R_b}\right )\left (\bar u_{L_c} t^ad_{R_d}\right )\right]+{\rm h.c.}\,,
\eea
where we used $C^{(1,8)\, abcd }_{quqd}=C^{(1,8)\, cbad }_{quqd}$, by virtue of Eq.\ \eqref{eq:MatchFQ}. The effective couplings of the four-quark operators are again proportional to $Z_{ab}$ for the terms where only a single flavor of up-type and down-type quarks appears.
In addition, the neutral-current pieces of the semileptonic operators, which give rise to EDMs, have the same couplings before and
after EWSB, due to our choice of flavor basis. The charged-current pieces are rotated by the CKM matrix and the PMNS matrix, $U$.
Both the semileptonic and four-quark operators \cite{Jenkins:2017dyc} follow the same RGEs as in Eq.\ \eqref{eq:RGE4fermion}.

The form of the dipole operators is slightly altered after EWSB as well, and we obtain a contribution to the Weinberg operator. This part of the CP-odd Lagrangian can be written as
\bea
\vL_{\rm dipole}&=&\sum_{l=e,\mu,\tau}\frac{-i e Q_l m_l}{2}C_l^{(\gamma)}\, \,\bar l\,\sigma^{\mu\nu}\gamma^5\,l\,F_{\mu \nu} +  \sum_{q=u,d,s,c,b}\frac{-i e Q_q m_q}{2}C_q^{(\gamma)}\, \,\bar q\,\sigma^{\mu\nu}\gamma^5\,q\,F_{\mu \nu}\nn\\
&&+  \sum_{q=u,d,s,c,b}\frac{-i g_s  m_q}{2}C_q^{(g)}\, \,\bar q\,\sigma^{\mu\nu}\gamma^5 t^a\,q\,G_{\mu \nu}^a+C_{\tilde
G}\frac{g_s}{6}f_{abc}\epsilon^{\al\bt\mu\nu}G^a_{\al\bt}G_{\mu\rho}^bG^{c\,\rho}_{\nu}\,.
\label{eq:EFTdipoles}
\eea
In addition,  two effective gluon-electron operators are induced 
\bea\label{eG}
\vL_{eG} = C_{eG}\,\al_s\bar e \, i\g_5 e \, G^a_{\mu\nu}G^{a\, \mu\nu}+ C_{e\tilde G}\,\frac{\al_s}{2}\bar e  e \, G^a_{\mu\nu}G^{a}_{\al\bt} \epsilon^{\al\bt\mu\nu}\,.
\eea
While these are operators of dimension 7, they can in principle be important for the charm and beauty quarks, since the additional
mass suppression is not severe compared to their sizable matrix element $\sim m_N$, the nucleon mass.

 For the matching of the four-fermion and dipole operators we simply have
\bea
C_{lequ}^{(1,3)}(m_t^-)&=&C_{lequ}^{(1,3)}(m_t^+)\,, \qquad C^{(1,8)}_{quqd}(m_t^-)=C_{quqd}^{(1,8)}(m_t^+)\,,\nn\\
C_{f}^{(\g)}(m_t^-)&=&C_{f}^{(\g)}(m_t^+)\,, \qquad C_{f}^{(g)}(m_t^-)=C_{f}^{(g)}(m_t^+)\,. 
\eea
In principle the four-fermion operators involving the top quark (or other heavy fermions) also give a matching contribution to the dipole operators. However, these contributions are proportional to $\ln(\mu/m_f)$ and thus vanish at the matching scale $\mu=m_f$.

The Weinberg operator obtains a contribution after integrating out the top quark
 \cite{BraatenPRL,Boyd:1990bx}
\bea
C_{\tilde G}(m_q^-) &=& C_{\tilde G}(m_q^+)-\frac{\al_s}{8\pi}C_q^{(g)}(m_q^+)\,.
\eea
Similar matching conditions apply at the bottom and charm thresholds, i.e.\ $q=c,b,t$. In principle,  the $C_{quqd}^{(1,8)\, q'qqq'}$ or $C_{quqd}^{(1,8)\, qq'q'q}$ couplings generate dimension-seven operators of the form $\frac{1}{m_{q'}}\bar q q \,GG$ after integrating out the heavier quark $q'$. These interactions in turn induce  the Weinberg operator after integrating out the lighter quark, $q$, which affects the couplings involving two heavy quarks, i.e.\ $Z_{cb,tb}$.  Here we only take into account the contributions from $Z_{cb,tb}$ to the Weinberg operator through the quark CEDMs. While both types of contributions appear at the two-loop level,  the former is suppressed by the lighter quark mass, $\sim m_q/m_{q'}$, while the latter is enhanced by the inverse ratio $\sim m_{q'}/m_q$.

Finally, the electron-gluon interactions are induced by one-loop diagrams involving the semileptonic four-fermion operators
\bea
C_{eG}(m_q^-) & =& C_{eG}(m_q^+) +\frac{1}{24\pi}\frac{{\rm Im}\, C_{le qu}^{(1)\, eeqq}(m_q^+)}{m_q}\,,\nn\\
C_{e\tilde G}(m_q^-) & =& C_{e\tilde G}(m_q^+) -\frac{1}{16\pi}\frac{{\rm Im}\, C_{le qu}^{(1)\, eeqq}(m_q^+)}{m_q}
\,.\eea
These contributions  only appear at the charm and top thresholds, such that $q=c,t$ here.

There are some more comments in order regarding the charm quark. Ideally, one would not integrate it out, since its mass is close
to the cut-off  scale of Chiral Perturbation Theory of the order of $1$ GeV, and instead use lattice QCD results to match to chiral low-energy constants.  However, since not
all required matrix elements are known, we estimate most of these contributions by integrating out the charm quark at $\mu=m_c$ as
indicated above. The two methods can be compared in the case of the contribution of $C_{lequ}^{(1)\, ee cc}$ to $C_S^{(0)}$, see Eq.~\eqref{eq:matchCSP}, yielding the relation $\sigma_c\approx 2m_N/27$. 
The value obtained in a recent lattice QCD calculation \cite{Alexandrou:2017qyt} is compatible with this estimate. 
On the other hand, for the charm EDM there is no sizable contribution to any of the operators discussed above after integrating out the
charm quark.  Nevertheless, the nucleon EDMs are expected to be induced non-perturbatively by the charm EDM. We therefore take this
contribution into account by explicitly including the charm tensor charge as discussed in the next section.

\begin{figure}
\begin{center}
\includegraphics[scale =.72]{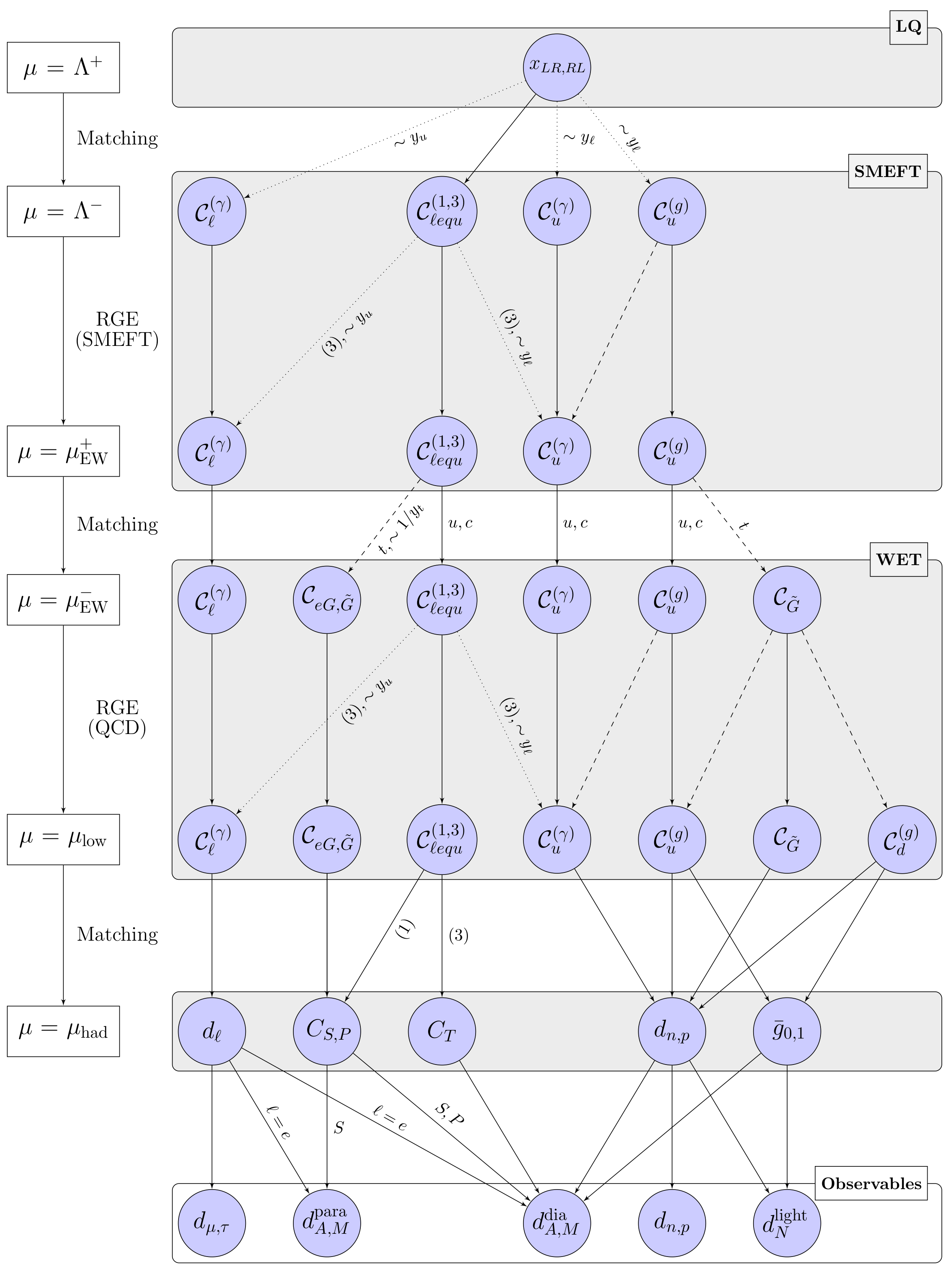}
\end{center}
\caption{Summary of the RGE and matching contributions induced by the $R_2$ or $S_1$ leptoquark interactions (with $x_{LR,RL}\to y_{LL,RR}$ for the latter) to the effective
CP-odd operators at lower energies, and finally to EDMs of various systems. WET stands for the weak effective theory below the
electroweak scale. $\mu_{\rm low}$ denotes a generic scale $\sim$GeV; in principle, there are additional matching steps at $\mu_b$
(which is negligible in this application) and $\mu_c$, discussed in the text. 
}
\label{Fig:RG}
\end{figure}

For the dipole and four-fermion operators the RGEs below $\mu=m_t$ are equivalent to those discussed in section \ref{RGE}, while the $C_{eG,e\tilde G}$ couplings do not run under QCD.
Thus, evaluating these RGEs and the matching contributions allows us to determine the couplings at low  energies, $\mu\simeq 1$ GeV.
The numerical results for the Wilson coefficients of relevance to EDMs are given in Table \ref{tab:RunningR2} for the $X_{lq} $
couplings, and their origins are depicted schematically in Fig.\ \ref{Fig:RG}. The pattern of $Y_{lq} $ couplings is essentially
identical because the RGEs are the same and the matching coefficients in Eqs.~\eqref{eq:MatchFQsemil} and \eqref{oneloopEDM1} are very
similar. The only exception are the coefficients of $X_{\ell t}$ in $C_\ell^{(\gamma)}$, which differ by almost a factor of 2, due
to different interference of the sizable direct matching contribution. We only show and discuss the $X_{lq} $ couplings in the
following. Table \ref{tab:RunningR2} only includes semileptonic interactions involving electrons, muons, and up quarks, while the interactions
involving heavier quarks and taus are integrated out.

For the leptoquark couplings involving electrons the low-energy Lagrangian is dominated by $C^{(\gamma)}_{e}$ and  $C_{{lequ}}^{(1,3)}$,
while purely hadronic contributions are suppressed with the electron mass and essentially negligible.
The couplings of electrons to the charm and top quarks induce a large $C^{(\gamma)}_{e}$. This is not surprising as the
one-loop contributions to the lepton EDMs in Eq.~\eqref{oneloopEDM1} scale with the quark mass and do not require an insertion
of the electron mass. 
As far as EDMs are concerned, the only low-energy difference between the $X_{e c}$ and $X_{e t}$ couplings is the relative size of
$C_{eG}$ which is significantly larger for $X_{e c}$. Nevertheless, even in this case the effect of $C_{eG}$ is suppressed compared to
$C^{(\gamma)}_{e}$, which will make it difficult to disentangle the electron-charm and electron-top leptoquark interactions.
On the other hand, for $X_{e u}$ the largest contribution is the CP-violating electron-quark coupling $C_{{lequ}}^{(1)}$ and this piece
will dominate atomic and molecular EDMs \cite{Fuyuto:2018scm}.

At first sight the couplings to muons show a similar pattern. The one-loop contributions to $C_\mu^{(\gamma)}$, the muon EDM, are
proportional to the mass of the quark running in the loop and thus grow for heavier quarks.
This means that  $C_\mu^{(\gamma)}$ is orders of magnitude larger than any other CP-odd low-energy interaction for $X_{\mu c}$ and
$X_{\mu t}$, while $C_{u}^{(\g,g)}$ is more important for $X_{\mu u}$. However, in these cases the smallness of the CP-violating
quark couplings can be misleading, because they contribute to hadronic and nuclear EDMs whose experimental limits are orders of
magnitude stronger than the limit on the muon EDM. For example, the limit on $X_{\mu c}$  will be still dominated by hadronic EDMs,
even though $C_\mu^{(\gamma)}$ is much larger than $C_{\tilde G}$.
For $X_{\mu t}$  the enhancement by the top mass
is sufficiently large that the muon EDM limit provides the strongest constraint.

For couplings to the tau, the story is similar to the muonic case. The main difference is that the enhancement of $C_\tau^{(\gamma)}$
with respect to $C_q^{(\gamma,q)}$  is smaller because of the larger tau mass. In addition, the experimental limit on the tau EDM is
weaker than that of the muon EDM by roughly two orders of magnitude. As such, the hadronic and nuclear EDMs tend to dominate the
experimental constraints for all tau couplings. However, taking the hadronic uncertainties into account, the $\tau$ EDM yields the
strongest constraint for $X_{\tau t}$.

Finally, the pattern of low-energy CP-odd operators for the $Z_{qq'}$ couplings is very different from that of the $X_{lq}$ couplings,
as indicated in Table~\ref{tab:RunningS1Quark}. The biggest change with respect to $X_{lq}$ and $Y_{lq}$ is that neither lepton EDMs
nor semileptonic four-fermion operators are generated. In general, the low-energy pattern is simple: if $a$ and $b$ are both light
quarks $(u,d,s)$ then CP-odd four-quark operators are generated at tree level and therefore in principle dominant.
If one quark is light and the other heavy (c,b,t), the dominant operators are the light-quark (chromo-)EDMs. If both quarks are
heavy, the biggest contribution is to the Weinberg operator with smaller contributions to light-quark (chromo-)EDMs.

\begin{table}
\small\center
$
\begin{array}{c|cccccccccccc}
R_2&X_{eu} &
   X_{ec} &  X_{et}&
   X_{\mu u} &  X_{\mu c}& X_{\mu
   t} &  X_{\tau u}& X_{\tau c}&
   X_{\tau t}\\\hline
  C_e^{(\gamma )} & 0.5 & 260 & 1\cdot 10^4 & 0 & 0 & 0 & 0 & 0 & 0 \\
 C_{\mu }^{(\gamma )} & 0 & 0 & 0 & 0.002 & 1 & 61 & 0 & 0 & 0 \\
 C_{\tau }^{(\gamma )} & 0 & 0 & 0 & 0 & 0 & 0 & 1\cdot 10^{-4} & 0.07 & \
4 \\
 C_u^{(\gamma )} &  \phantom{-}0.02 &0 & 0 & 5 & -4\cdot 10^{-9} & -6\cdot 10^{-10} & 70 & -6\
\cdot 10^{-8} & -1\cdot 10^{-8} \\
 C_u^{(g)} & -6\cdot 10^{-4} &0& 0 & \
-0.1 & -1\cdot 10^{-7} & -1\cdot 10^{-9} & -2 & -2\cdot 
10^{-6} & -2\cdot 10^{-8} \\
 C_{d,s}^{(\gamma )} & 0 &0& 0& 0 & -4\cdot 10^{-9} & -6\cdot 10^{-10} & 0 & -6\cdot 10^{-8} & -1\cdot 10^{-8} \\
 C_{d,s}^{(g)} & 0 &0 &0 & 0 & \
-1\cdot 10^{-7} & -1\cdot 10^{-9} & 0 & -2\cdot 10^{-6} & \
-2\cdot 10^{-8} \\
 C_c^{(\g)} &0 &4\cdot 10^{-5}& 0 & \
0 & \phantom{-\,}9\cdot 10^{-3} & 0 &0 &0.1 & 0 \\
 C_{\tilde{G}} & 0 &0 &0& 0 &  \phantom{-\,}3\cdot 10^{-6} &  \phantom{-\,}2\cdot 10^{-9} & 0 &  \phantom{-\,}4\cdot 10^{-5} & \phantom{-\,\,}3\cdot 10^{-8} \\
 C_{{lequ}}^{(1)\, lluu}& -1\dt^l_e & 0 & 0 & -1\dt^l_\mu & 0 & 0 & 0 & 0 & 0 \\
 C_{{lequ}}^{(3)\, lluu} & -0.1\dt^l_e & 0 & 0 &  -0.1\dt^l_\mu & 0 & 0 & 0 & 0 & 0 \\
 C_{{eG}} & 0 & -0.01 & -5\cdot 10^{-5} & 0 & 0 & 0 & 0 & 0 & 0 \\
 C_{e\tilde{G}} & 0 &  \phantom{-\,}0.02 &  \phantom{-\,}7\cdot 10^{-5} & 0 & 0 & 0 & 0 & 0 & 0 \\
\end{array}
$
\caption{Dependence of the low-energy CP-odd Wilson coefficients in Eqs.\ \eqref{eq:EFT4fermion}, \eqref{eq:EFTdipoles}, and \eqref{eG}
on the $X_{lq}$ couplings for  $m_{LQ}=1$ TeV. All low-energy couplings were evaluated at $\mu=1$ GeV apart from $C_c^{(\g)}$ which was
evaluated at $\mu=m_c$.}
\label{tab:RunningR2}
\end{table}

\begin{table}
\small\center
$
\begin{array}{c|ccccccccc}
S_1& Z_{ud} & Z_{us}& Z_{ub}& Z_{cd} & \
Z_{cs} & Z_{cb} & Z_{td} & Z_{ts} & Z_{tb} \\\hline
 C_u^{(\gamma )} & -0.07 & -1 & -52 & 0 & 0 & -2\cdot 10^{-5} & 0 & 0 & -6\cdot 10^{-4} \\
 C_u ^{(g)} & -0.5 & -10 & -380 &0 & 0& -4\cdot 10^{-4} &0 &0& \
-3\cdot 10^{-3} \\
 C_d ^{(\gamma )} &  \phantom{-}0.06 & 0 & 0 &  \phantom{-}31 &0& -2\cdot 10^{-5} &  \phantom{-}1600 & 0 & -6\cdot 10^{-4} \\
 C_d ^{(g)} & -0.10 & 0 &0 & -55 & 0& -4 \cdot 10^{-4} & -3000 &0& -3\cdot 10^{-3} \\
 C_s ^{(\gamma )} & 0 & \phantom{-}3\cdot 10^{-3} &0&0 &  \phantom{-}2 & -2\cdot 10^{-5} &0 &  \phantom{-}78 & -6 \cdot 10^{-4} \\
 C_s^{(g)} & 0 & -5\cdot 10^{-3} & 0&0
& -3 & -4\cdot 10^{-4} & 0& -150 & -3\cdot 10^{-3} \\
 C_c^{(\g)} &0 &0& 0 & -1 \cdot 10^{-4} &-2  \cdot 10^{-3} & -8 \cdot 10^{-2} &0 &0 & -4  \cdot 10^{-4} \\
 C_{\tilde{G}} & 0 & 0 & 0 & 0& \
0 & \phantom{-}8\cdot 10^{-3} & 0& 0 & 0.01 \\
 C_{{quqd}}^{(1),uuqq} & -4\delta^q_d & -4\delta^q_s & 0 & 0 & 0 & 0 & 0 & 0 & 0 \\
 C_{{quqd}}^{(8),uuqq} &  \phantom{-}4\delta^q_d & \phantom{-}4\delta^q_s & 0 & 0 & 0 & 0 & 0 & 0 & 0 
\end{array}
$
\caption{Dependence of the low-energy CP-odd Wilson coefficients in Eqs.\ \eqref{eq:EFT4fermion} and \eqref{eq:EFTdipoles}
on the $Z_{qq'}$ couplings for  $m_{LQ}=1$ TeV. All low-energy couplings were evaluated at $\mu=1$ GeV apart from $C_c^{(\g)}$ which
was evaluated at $\mu=m_c$.}
\label{tab:RunningS1Quark}
\end{table}

\section{Matching to even lower energies}\label{sec:4}
Apart from the muon and tau EDM operators none of the CP-violating operators in Eqs.~\eqref{eq:EFT4fermion}, \eqref{eq:EFTdipoles}, and
\eqref{eG} are measured directly. Most EDM experiments involve complex objects like nucleons, nuclei, atoms, and molecules. The 
challenge is to connect the operators at the partonic
level to the observables measured in the laboratory. To do so, it has proven useful to first match to an effective description in terms
of the low-energy degrees of freedom: pions, nucleons, leptons, and photons. This matching can be systematically performed by using
(baryon) chiral perturbation theory ($\chi$PT), the low-energy EFT of QCD, extended to include CP violation
\cite{deVries:2012ab}.
The big advantage of $\chi$PT is that observables can be calculated in perturbation theory in an expansion in $q/\Lambda_\chi$ where $q$ is the
typical momentum scale in the observable and $\Lambda_\chi \simeq 1$ GeV the chiral-symmetry-breaking scale. $\chi$PT makes it possible
to construct the effective hadronic interactions order by order in perturbation theory which can then be used to calculate nucleon and
nuclear EDMs.

Detailed studies \cite{deVries:2012ab, Engel:2013lsa,Yamanaka:2017mef} show that EDMs of current experimental interest can be calculated at leading order in terms of a handful effective CP-odd interactions. The first set of operators relevant at low energies are trivial and consist of the lepton EDMs, which are given by 
\bea
\mathcal L_{\mathrm{LEDM}} &=&  \sum_{l=e,\mu,\tau}-\frac{d_l}{2}\, \,\bar l\,i\sigma^{\mu\nu}\gamma^5\,l\,F_{\mu \nu}\,,\qquad d_l = e\, Q_l m_l C_{l}^{(\g)}\,.
\label{eq:leptonEDMs}\eea
At this point the $C_{lequ}^{(3)\, lluu}$ operators, relevant for the $X_{lu}$ and $Y_{lu}$ couplings, are still present in the EFT and
can also contribute to the lepton EDMs. These contributions depend on an unknown hadronic matrix element, but a
naive-dimensional-analysis  (NDA) estimate \cite{Weinberg:1989dx,NDA} would predict $d_l \sim\frac{F_\pi}{4\pi} C_{lequ}^{(3)\, lluu}$,
potentially making this the dominant contribution \cite{Dekens:2018pbu}. Nevertheless, for $\ell=e$ the $C_{lequ}^{(1,3)}$ operators can contribute directly
to atomic EDMs which leads to more important effects than $d_e$,  while for $\ell=\mu,\tau$ the contributions to $C_u^{(\gamma,g)}$ are
far more relevant.  We  therefore neglect any $C_{lequ}^{(3)\, lluu}$ contributions in Eq.~\eqref{eq:leptonEDMs}.

The remaining relevant leading-order operators contain hadrons and are given by\footnote{We have written the interactions in terms of
non-relativistic heavy-nucleon fields, appropriate for hadronic and nuclear studies of CP violation. The pseudoscalar and tensor
semileptonic CP-violating interactions often appear in the literature  as 
$\mathcal L =-\frac{G_F}{\sqrt{2}}
\bigg\{\bar e e\,\bar
\Psi_N \left(C_P^{(0)}+\tau_3 C_P^{(1)}\right)i \gamma^5  \Psi_N 
+ \bar e \sigma^{\mu\nu}e\,\bar \Psi_N \left(C_T^{(0)}+\tau_3 C_T^{(1)}\right)
 i\sigma_{\mu\nu}\gamma_5 \Psi_N\bigg\}$ in terms of relativistic nucleon fields
$\Psi_N$.} 
\bea\label{hadronicCPV} \vL & =& \bar g_0 \bar N \boldtau\cdot \boldpi N +\bar g_1 \bar N \pi_3 N - 2 \bar N(\bar
d_0+\bar d_1 \tau_3)S^\mu N\,v^\nu F_{\mu\nu} \nn\\
&&-\frac{G_F}{\sqrt{2}}\bigg\{\bar e i\g_5 e\, \bar N\left(C_S^{(0)}+\tau_3 C_S^{(1)}\right) N + \bar e e\, \frac{\partial_\mu}{m_N} \left[\bar N\left(C_P^{(0)}+\tau_3 C_P^{(1)}\right)S^\mu N\right]\nn\\
&&-4\, \bar e \sigma_{\mu\nu} e\, \bar N\left(C_T^{(0)}+\tau_3 C_T^{(1)}\right) v^\mu S^\nu N\bigg\}+\dots \,,
\eea
in terms of the Pauli matrices $\boldtau$, the electron field $e$, the pion triplet $\boldpi$, the non-relativistic nucleon doublet $N
= (p\,\,n)^T$ and its mass $m_N$, the velocity $v^\mu$, and the spin $S^\mu$ ($v^\mu =(1,\boldsymbol 0)$ and $S^\mu = (0,\, \boldsymbol
\sigma/2 )$ in the nucleon rest frame). The dots denote additional interactions that in principle appear at leading order, such as
CP-odd nucleon-nucleon interactions, but were found to lead to small contributions in explicit calculations on light nuclei \cite{deVries2011b,Bsaisou:2014zwa}. The
coupling constants, usually called low-energy constants (LECs), $\bar g_{0,1}$, $\bar d_{0,3}$, and $C^{(0,1)}_{S,P,T}$ cannot be
obtained from symmetry arguments alone and need to be fitted to data or obtained in a non-perturbative calculation, for instance via
lattice QCD methods.

We begin with discussing the CP-odd pion-nucleon LECs $\bar g_{0,1}$. These interactions involve non-derivative pion couplings and are
only effectively induced by CP-odd sources that violate chiral symmetry, which, in the leptoquark context are the quark chromo-EDMs and the four-quark operators involving strangeness (the four-quark interactions without strange quarks are chirally invariant, while quark EDM operators do violate chiral symmetry but contain an explicit photon which needs to be integrated out to
induce $\bar g_{0,1}$, such that the resulting contributions are suppressed by $\alpha_{\mathrm{em}}/\pi$). The exact sizes of $\bar
g_{0,1}$ are not well known but a QCD sum rules calculation gives \cite{Pospelov_piN} 
\bea
\bar g_0 &=& (5\pm 10)(m_u\tilde C^{(u)}_g + m_d\tilde C^{(d)}_g)\, \mathrm{fm}^{-1}
+\frac{\Lambda_\chi^2}{4\pi}{\rm Im}\left(C_{quqd}^{(1)\, a ss u}V_{ua}^*\right)+\frac{\Lambda_\chi^2}{4\pi}{\rm Im}\left(C_{quqd}^{(8)\, a ss u}V_{ua}^*\right)\nn
\,,\\
\bar g_1 &=& (20^{+40}_{-10})(m_u\tilde C^{(u)}_g- m_d\tilde C^{(d)}_g)\,\, \mathrm{fm}^{-1}
+\frac{\Lambda_\chi^2}{4\pi}{\rm Im}\left(C_{quqd}^{(1)\, a ss u}V_{ua}^*\right)+\frac{\Lambda_\chi^2}{4\pi}{\rm Im}\left(C_{quqd}^{(8)\, a ss u}V_{ua}^*\right)
\eea
where we used NDA estimates for the four-quark contributions, to which we assign a $90\%$ uncertainty, i.e.\ we use $\Lambda_\chi^2/(4\pi)(1\pm0.9)$.\footnote{It should be mentioned that part of the contribution to $\bar g_{0,1}$ can be extracted from lattice calculations for different operators with a similar chiral structure $\sim \bar u_L\bar u_R \,\bar s_L s_R$ \cite{Cirigliano:2016yhc}. In this case these contributions exceed the NDA expectation by roughly an order of magnitude.}
Contributions from the strange CEDM are suppressed by the small $\eta$-$\pi$ mixing angle \cite{deVries:2016jox}, while those from the Weinberg operator and  $C^{(1,8)}_{quqd}$ appear at next-to-next-to-leading order in the chiral expansion \cite{deVries:2012ab}.

We now turn to the EDMs of the neutron, $d_n = \bar d_0 - \bar d_1$, and proton, $d_p = \bar d_0 + \bar d_1$, which are induced by
quark (color-)EDMs, the four-quark interactions, and the Weinberg operator. Because of the many contributions the expressions are
lengthy,
\bea
d_n&=&
g_T^u\,d_u^{\rm eff}+g_T^d\,d_d+g_T^s\,d_s+g_T^c\,d_c\nn\\
&&-(0.55\pm0.28)\,e\,\tilde d_u-(1.1\pm0.55)\,e\,\tilde d_d \pm(50\pm40)\,{\rm MeV}\,e\,g_sC_{\tilde G}\nn\\
&&\pm (11\pm10)\,{\rm MeV}\,e\, {\rm Im}\left(C_{quqd}^{(1)\, a dd u}V_{ua}^*\right)\pm (11\pm10)\,{\rm MeV}\, \,e\,{\rm Im}\left(C_{quqd}^{(8)\, a dd u}V_{ua}^*\right)\nn\\
&&\pm (11\pm10)\,{\rm MeV}\,e\,{\rm Im}\left(C_{quqd}^{(1)\, a ss u}V_{ua}^*\right)\pm (11\pm10)\,{\rm MeV}\,e\,{\rm Im}\left(C_{quqd}^{(8)\, a ss u}V_{ua}^*\right)\,,\nn\\
d_p&=&
g_T^d\,d_u^{\rm eff}+g_T^u\,d_d+g_T^s\,d_s+g_T^c\,d_c\nn\\
&&+(1.30\pm0.65)\,e\, \tilde d_u+(0.60\pm0.30)\,e\,\tilde d_d \mp(50\pm40)\,{\rm MeV}\,e\,g_sC_{\tilde G}\nn\\
&&\mp (11\pm10)\,{\rm MeV}\, \,e\, {\rm Im}\left(C_{quqd}^{(1)\, a dd u}V_{ua}^*\right)\mp (11\pm10)\,{\rm MeV}\, e\,{\rm Im}\left(C_{quqd}^{(8)\, a dd u}V_{ua}^*\right)\nn\\
&&\pm (11\pm10)\,{\rm MeV}\,e\,{\rm Im}\left(C_{quqd}^{(1)\, a ss u}V_{ua}^*\right)\pm (11\pm10)\,{\rm MeV}\,e\,{\rm Im}\left(C_{quqd}^{(8)\, a ss u}V_{ua}^*\right)\,,
\eea
where all coefficients should be evaluated at $\mu=\Lambda_\chi=1$ GeV apart from the explicit charm contribution, where
$\mu=2$~GeV, and
\bea 
d_u^{\rm eff} = d_u (\Lambda_\chi)+e \sum_{l=e,\mu}Q_lm_l \frac{16}{(4\pi)\sq}C_{lequ}^{(3)\, lluu}(\Lambda_\chi)\ln(\Lambda_\chi/m_l)\,,\label{eq:nucleonEDMs}
\eea
where the second term arises from one loop diagrams involving the semileptonic operators, $C_{lequ}^{(3)}$, which effectively induce
the up-quark EDM.\footnote{The four-quark operators involving light quarks in principle give rise to similar contributions to the `effective' quark (C)EDMs. However, in these cases we simply absorb these terms into the sizable theoretical uncertainties of the matrix elements in the last lines of Eqs~\
\eqref{eq:nucleonEDMs}.}

The contributions from the first-generation quark EDMs to $d_{n,p}$ are known to $\Or(5\%)$ from   lattice calculations \cite{Bhattacharya:2015esa, Bhattacharya:2015wna,Bhattacharya:2016zcn,Gupta:2018qil,Gupta:2018lvp}, while the strange contribution is smaller and has a larger relative uncertainty. At $\mu=1$ GeV we have,
\bea
 g_T^u &=& -0.213 \pm 0.011\,,\qquad g_T^d = 0.820 \pm 0.029\,,\qquad g_T^s =- 0.0028\pm 0.0017\,.
\eea
The charm contribution is even smaller and its value consistent with zero so far \cite{Alexandrou:2017qyt}
\begin{equation}\label{charmTensor}
g_T^c = -0.0027\pm0.0028\quad(\mu=m_c)\,.
\end{equation}
Since this contribution yields potentially the strongest limits on two of the
phenomenologically important leptoquark couplings (see Sect.~\ref{pheno}), an improved estimate of this matrix element would be very
welcome. The central value of the charm tensor charge is presently comparable to the strange tensor charge which is surprising and we
expect the actual matrix element to be smaller. In what follows below we will present two types of limits on leptoquark interactions
based on different ways of handling the theoretical uncertainty in the matrix element. In the `Central' strategy we typically take the
central value of the matrix elements as given in this section. For the charm tensor charge, however, we use $g_T^c \rightarrow
(m_s/m_c)g_T^s \simeq 0.08\,g_T^s \simeq -2.2\cdot 10^{-4}$ for the central value to account for the expected relative suppression of charm contributions to
the nucleon EDMs.

The contributions from the up and down quark CEDMs have been estimated using QCD sum-rule calculations \cite{Pospelov_qCEDM,
Pospelov_deuteron, Pospelov_review, Hisano1}. The contributions from the strange CEDM are usually considered to vanish, once a
Peccei-Quinn mechanism is used to solve the strong CP problem \cite{Pospelov_qCEDM}, but this has not been fully resolved
\cite{Hisano2}. Contributions from the Weinberg operator appear with large uncertainties, $\Or(100\%)$, based on a combination of QCD
sum-rules \cite{Pospelov_Weinberg} and naive-dimensional-analysis, 
Lattice-QCD calculations are
in progress to reduce the uncertainties \cite{Abramczyk:2017oxr,Dragos:2017wms,Yoon:2017tag}.
Very little is known about the contributions of the four-quark operators and we use the NDA estimate $d_{n,p}=\Or(\Lambda_\chi/(4\pi)\sq)\, {\rm Im}\,C_{quqd}^{(1,8)}$ with an $\Or(100\%)$ uncertainty \cite{deVries:2012ab}.

Finally, we discuss the electron-nucleon interactions that are induced by $C_{lequ}^{(1,3)}$ and $C_{eG,e\tilde G}$: 
\bea \label{eq:matchCSP}
C_S^{(0)} &=& v\sq\left[\frac{\sigma_{\pi N}}{m_u+m_d}{\rm Im}\, C_{lequ}^{(1)\, eeuu}+\frac{16\pi}{9}(m_N-\sigma_{\pi N}-\sigma_s)C_{eG}\right]\,,\nn\\
C_S^{(1)}& =& v\sq\frac{1}{2}\frac{\dt m_N}{m_d-m_u}{\rm Im}\, C_{lequ}^{(1)\, eeuu}\,,\nn\\
C_P^{(0)} &=&-8\pi v^2(\Delta_u+\Delta_d) m_NC_{e\tilde G}\,,\qquad C_P^{(1)} = v\sq \frac{g_Am_N}{m_u+m_d}{\rm Im}\, C_{lequ}^{(1)}-8\pi v^2g_A m_N\frac{m_d-m_u}{m_u+m_d}C_{e\tilde G} \,,\nn\\
C_T^{(0)} &=& v^2(g_T^d+g_T^u){\rm Im}\, C_{lequ}^{(3)\, eeuu}\,,\qquad\,\,\, C_T^{(1)} \,\,= v^2(g_T^d-g_T^u){\rm Im}\, C_{lequ}^{(3)\, eeuu}\,.
\eea
The  hadronic matrix elements needed for the contributions to $C_S^{(0,1)}$ are the  scalar charges of the nucleons, which are related to the nucleon sigma terms, $\sigma_{\pi N, s}$, and the strong part of the  nucleon mass splitting, $\dt m_N = (m_n-m_p)^{\rm QCD}$. Instead, the contributions to the $C_T^{(0,1)}$ interactions depend on the nucleon tensor charges, $g_{T}^{u,d}$. Finally, the contributions to $C_P^{(0,1)}$ depend on the isoscalar and isovector axial charges,  $\Delta_u+\Delta_d$ and $g_A =  \Delta_u - \Delta_d$, respectively\footnote{The contribution of $C_{lequ}^{(1)}$ to $C_P^{(1)}$ in Eq.\ \eqref{eq:matchCSP} arises through the exchange of a pion, the so-called pion pole, where we have approximated $m_\pi\sq/(q\sq-m_\pi\sq)\simeq -1$. To obtain the contributions from $C_{e\tilde G}$ we used an $U(1)_A$ rotation to rewrite $G_{\mu\nu}\tilde G^{\mu\nu}$ in terms of $\partial_\mu \bar q\g^\mu \g_5 q$ and $\bar qi \g_5 q$. The hadronization of these terms then leads to the appearance of the axial charges.}.
The relevant hadronic input for the axial charges \cite{Airapetian:2006vy}, $\sigma_{\pi N}$ \cite{Hoferichter:2015dsa}, $\sigma_s$ \cite{Abdel-Rehim:2016won} and $\dt m_N$ \cite{Borsanyi:2014jba,Brantley:2016our} can be summarized as 
\bea
\sigma_{\pi N} &=& (59.1 \pm 3.5)\,\mathrm{MeV}\ ,\qquad \sigma_s = (41.1_{-10.0}^{+11.3})\,\mathrm{MeV} \, ,\qquad 
\delta m_N =( 2.32\pm0.17)\, \mathrm{MeV}\, ,\nn\\
  g_A &=& 1.27\pm0.002\,,\qquad \Delta_u = 0.842\pm 0.012\,,\qquad \Delta_d = -0.427\pm 0.013\,.
\eea
Recent lattice-QCD calculations typically find smaller values for $\sigma_{\pi N}$ \cite{Durr:2015dna, Bali:2016lvx, Yamanaka:2018uud}.

Of the hadronic CP-odd interactions in Eq.~\eqref{hadronicCPV} only the neutron EDM is measured directly. The proton EDM could
potentially be probed directly in a future electromagnetic storage ring \cite{Anastassopoulos:2015ura}. Connecting most of
the interactions in Eq.~\eqref{hadronicCPV} to actual EDM measurements therefore requires one further step.

\subsection{Contributions to nuclear, atomic, and molecular EDMs}

Currently, the strongest experimental limit is set on the EDM of the $^{199}$Hg atom. This is a diamagnetic system and therefore 
no large enhancement factors mitigate the Schiff screening by the electron cloud \cite{Schiff:1963zz}. The main contributions are
hence expected from the nuclear Schiff moment and semileptonic interactions. The $\bar g_{0,1}$ contributions entering the
expression for the Schiff moment require complicated many-body calculations which at present cannot be performed with good theoretical
control \cite{Dmitriev:2003sc,deJesus:2005nb,Ban:2010ea,Dzuba:2009kn,Engel:2013lsa}, leading to large nuclear uncertainties. For the
(semi-)leptonic contributions the calculations are under much better control, see
Refs.~\cite{Yamanaka:2017mef,Singh:2014jca,Sahoo:2016zvr,2018arXiv180107045S,Fleig:2018bsf} for recent results.

Collecting all the different contributions, we obtain
\cite{Engel:2013lsa,Singh:2014jca,Sahoo:2016zvr,2018arXiv180107045S,1402-4896-36-3-011,Fleig:2018bsf,Dzuba:2009kn,Latha:2009nq,Radziute:2015apa,Yamanaka:2017mef,Dmitriev:2003sc}
\bea\label{dHg} d_{\rm Hg}&=& -(2.1\pm0.5)
\Ex{-4}\bigg[(1.9\pm0.1)d_n +(0.20\pm 0.06)d_p+\bigg(0.13^{+0.5}_{-0.07}\,\bar g_0 +
0.25^{+0.89}_{-0.63}\,\bar g_1\bigg)e\, {\rm fm}\bigg]\nn\\
&&+(0.012\pm0.012) d_e - \left[ (0.028\pm0.006) C_S- \frac{1}{3}(3.6\pm0.4) \left(C_T+\frac{Z\al}{5 m_N R}C_P\right)\right]\cdot
10^{-20}\, e\,\mathrm{cm}\,,\nn\\
\eea
where $C_{S,P}$ and $C_T$ are effective scalar and tensor couplings.
The effective scalar coupling depends on the numbers of protons ($Z$) and neutrons ($N$),  $C_S = C_S^{(0)}+\frac{Z-N}{Z+N} C_S^{(1)}$.
While this renders $C_S$ in principle system-dependent, it turns out that the variation for the heavy systems under consideration
is negligible and hence the same coefficient can be used for Hg and  all paramagnetic
systems discussed below \cite{Jung:2013mg,Jung:2013hka}.
The same is not true for the pseudoscalar and tensor matrix elements:
they are related
\cite{Dzuba:2009kn} and we have $C_{P,T} = (C_{P,T}^{(n)}\langle \vec \sigma_n\rangle  +C_{P,T}^{(p)}\langle \vec \sigma_p\rangle
)/(\langle \vec \sigma_n\rangle +\langle \vec \sigma_p\rangle )$, with $C_{P,T}^{(n,p)}=C_{P,T}^{(0)}\mp C_{P,T}^{(1)}$. For $^{199}$Hg
we have \cite{Yanase:2018qqq} 
\bea
\langle \vec \sigma_n\rangle= -0.3249\pm0.0515\,,\qquad \langle \vec \sigma_p\rangle = 0.0031\pm 0.0118\,,
\eea
 so that $C_{P,T}\simeq C_{P,T}^{(0)}-C_{P,T}^{(1)}$ \cite{Jung:2013hka}. $R\simeq 1.2\, A^{1/3}$ fm is the nuclear radius in terms of $A=Z+N$. 

The number of terms in Eq.~\eqref{dHg} shows the necessity to measure the EDMs of as many different diamagnetic systems as
possible in order to disentangle the various contributions. At present no other EDM measurement of a diamagnetic system comes close to the precision of the $^{199}$Hg measurement.
However, experimental efforts are ongoing to measure for instance the EDMs of ${}^{129}$Xe
\cite{2002PhLA..304...13Y,2005PhRvL..94f0801L,Heil:2013tpa,Kuchler:2014gda}, the diamagnetic molecule TlF
\cite{2012PhRvA..85a2511H,PhysRevA.95.062506}, and ${}^{225}$Ra  (to improve the recent results in Ref.~\cite{Parker:2015yka,Bishof:2016uqx}), 
each aiming at improving existing limits by several orders of magnitude. These measurements are essential to obtain model-independent
information from diamagnetic systems, even if a given measurement might not give the best limit on an individual coupling.

We include ${}^{225}$Ra exemplarily for these new efforts, whose EDM limit \cite{Bishof:2016uqx} is currently  about six orders
of magnitude weaker than the $^{199}$Hg limit. Nevertheless, this is an interesting system because of the octopole deformation of its
nucleus which greatly enhances the contribution from the CP-odd pion-nucleon couplings. Neglecting all other (smaller) contributions we
write \cite{Engel:2013lsa,Dobaczewski:2018nim}~\footnote{
The neglected contributions are not expected to exhibit the octupole enhancement, as can be seen for semileptonic
contributions explicitly, see Ref.~\cite{Singh:2015aba} and references therein for recent calculations. The limits on such
contributions arising from present or even presently projected measurements are hence not expected to be competitive with those obtained from mercury.}
\bea\label{dRa} d_{\mathrm{Ra}} &=& (-7.7\pm
0.8)\Ex{-4}\cdot\left[\left(-2.5\pm 7.6\right) \,\bar g_0 + \left(63\pm 38\right)\,\bar g_1\right]e\, {\rm fm}\,.
\eea
Despite the large nuclear coefficients, the current limit is not competitive.
Ongoing efforts aim to reach a sensitivity $d_{\mathrm{Ra}} < 10^{-27}\,e\,\mathrm{cm}$.

The EDM in heavy paramagnetic systems is characterized by large enhancement factors for the electron EDM and the scalar
electron-nucleon coupling $C_S$. The best available limit from an atom stems from Thallium, whose EDM can be expressed as
\cite{Dzuba:2009mw,Porsev:2012zx}
\bea
d_{\rm Tl} &=& (-573\pm 20) d_e - (700 \pm 35)\cdot 10^{-20}\, e\,\mathrm{cm}\, C_S\,.
\eea
Currently, measurements of molecular systems give rise to the most stringent constraints on the electron EDM and
electron-nucleon couplings, due to the huge effective inner-molecular electric field. We use 
\cite{doi:10.1063/1.4968597,Skripnikov,PhysRevA.96.040502,doi:10.1063/1.4968229,PhysRevA.90.022501,PhysRevA.93.042507}
\bea
\omega_{\text{YbF}} &=& (-19.6\pm1.5)(\mathrm{mrad}/\mathrm{s})\left(\frac{d_e}{10^{-27}\,e\,\mathrm{cm}}\right)-(17.6\pm2.0)(\mathrm{mrad}/\mathrm{s})\left(\frac{C_S }{10^{-7}}\right)\,,\\
\omega_{\text{HfF}} &=&(34.9\pm1.4)(\mathrm{mrad}/\mathrm{s})\left(\frac{d_e}{10^{-27}\,e\,\mathrm{cm}}\right)+(32.0\pm1.3)(\mathrm{mrad}/\mathrm{s})\left(\frac{C_S }{10^{-7}}\right)\,,\\
\omega_{\text{ThO}} &=&(120.6\pm4.9)(\mathrm{mrad}/\mathrm{s})\left(\frac{d_e}{10^{-27}\,e\,\mathrm{cm}}\right)+(181.6\pm7.3)(\mathrm{mrad}/\mathrm{s})\left(\frac{C_S }{10^{-7}}\right)\, .
\eea
$C_S$ is defined below Eq.~\eqref{dHg}; in that expression $Z$ and $N$ of the heaviest atom of the molecule should be used,
yielding an approximately universal coefficient. There are various experimental efforts underway, as an illustration we use an
expected improved limit on ThO.

So far no experimental limits have been set on the EDMs of nuclei although advanced proposals exist to measure the EDMs of light nuclei
in electromagnetic storage rings \cite{Eversmann:2015jnk}. In this work we consider the impact of a direct measurement of the deuteron
EDM which can be accurately expressed \cite{Bsaisou:2014zwa,Yamanaka:2015qfa} in terms of the interactions in Eq.~\eqref{hadronicCPV}:
\begin{eqnarray}
d_{D} &=&
(0.94\pm0.01)(d_n + d_p) + \bigl [ (0.18 \pm 0.02) \,\bar g_1\bigr] \,e \,{\rm fm} \, .
 \label{eq:h2edm} 
\end{eqnarray}

\begin{table}[t]
\small\center
\renewcommand{\arraystretch}{1.2}
$\begin{array}{ccccc|ccc}
\multicolumn{5}{c|}{{\rm Particles, hadrons,}\, {\rm and }\,{\rm atoms }  \,(e \, {\rm cm})} &\multicolumn{3}{c}{{\rm Molecules}  \,(\mathrm{mrad}/\mathrm{s}) }   \\\hline
d_\mu& d_\tau &d_n & d_{\rm Hg} & d_{\rm Tl} & \omega_{\text{YbF}}&\omega_{\text{HfF}} & \omega_{\text{ThO}}\\\hline
1.5\cdot 10^{-19}&3.4\cdot 10^{-17}& 
3.0 \cdot 10^{-26} &6.3\cdot 10^{-30} & 9.4\cdot 10^{-25}&
23.5
&  4.6
&
1.3\\
  \end{array}$
\caption{Current experimental limits (at 90$\%$ C.L.) from measurements on the muon \cite{Bennett:2008dy}, tau \cite{Inami:2002ah},
neutron \cite{Afach:2015sja,Baker:2006ts}, $^{199}$Hg \cite{Griffith:2009zz,Graner:2016ses}, Tl \cite{Regan:2002ta}, YbF
\cite{Hudson:2011zz,Kara:2012ay}, HfF \cite{Cairncross:2017fip}, and ThO \cite{Baron:2013eja,Baron:2016obh,Andreev:2018ayy}.}\label{tab:expt}
\end{table}

\begin{table}[t]
\small\center
\renewcommand{\arraystretch}{1.2}
$\begin{array}{c|ccccc|c}
\multicolumn{1}{c}{}  &\multicolumn{5}{c|}{{\rm Particles, hadrons, nuclei,}\, {\rm and }\,{\rm atoms }  \,(e \, {\rm cm})} &\multicolumn{1}{c}{{\rm Molecules}  \,(\mathrm{mrad}/\mathrm{s}) }   \\ \hline
&d_\mu &d_\tau & d_n &d_{p,D}   & d_{\rm Ra} & \omega_{\text{ThO}} \\\hline
\mathrm{current} &1.5 \cdot 10^{-19}  & 3.4\cdot 10^{-17}  & 3.0 \cdot 10^{-26}  & -    & 1.2\cdot 10^{-23}  & 1.3\\
\mathrm{expected} &1.0 \cdot 10^{-21}    & 6 \cdot 10^{-19}  &1.0 \cdot 10^{-28}& 1.0 \cdot 10^{-29} & 1.0 \cdot 10^{-27} &0.1\
  \end{array}$
\caption{Expected sensitivities of several promising future EDM experiments, see Refs.~\cite{Kou:2018nap,Chen:2018cxt, Chupp:2017rkp, Holzbauer:2017ntd}.}\label{tab:exptfut}
\end{table}

\section{Constraints on leptoquark interactions}\label{pheno}
We now discuss the limits that can be set on the various CP-violating combinations of leptoquark couplings defined in
Eq.~\eqref{defLQ}, using the current and projected experimental EDM limits in Tables~\ref{tab:expt} and  \ref{tab:exptfut}. Here we
also consider the proton, deuteron and radium EDMs, for which current limits do not play a role, but prospected sensitivities would
lead to impressive improvements. 
We define a $\chi^2$ function in the standard way
\begin{equation}
\chi_i^2 = \left(\frac{\mathcal O_i^{th}-\mathcal O_i^{exp}}{\sigma_i}\right)^2\,,
\end{equation}
where $\mathcal O_i^{exp}$ stands for the experimentally measured value of a particular EDM (these measurements are all
null-measurements at present), $\mathcal O_i^{th}$ is the theoretical expression given above, and $\sigma_i$ is the experimental
uncertainty. 
In addition, we have to decide how to handle the theoretical uncertainties in the hadronic, nuclear, and atomic matrix elements that
connect the EDM limits to the fundamental CP-violating couplings. In several cases, these matrix elements have large uncertainties. For
instance, the coefficients linking the ${}^{199}$Hg EDM to the CP-odd pion-nucleon couplings span a large range that sometimes even
includes zero. In order to understand the role of these theoretical uncertainties, we adopt to two very different strategies:   
\begin{itemize}
\item \textbf{Central:} This is the ``optimistic" strategy where all theoretical uncertainties are neglected and we simply take the
central values of all hadronic, nuclear, and atomic matrix elements. Its purpose is twofold: it shows the general
sensitivity of the observable in question to a specific source and also illustrates what could be achieved with present data for
future improved theory calculations of the matrix elements. This strategy correspondingly leads to rather strong constraints.
\item \textbf{R-fit:} This is the ``pessimistic" strategy, where we vary all matrix elements within their allowed theoretical ranges
discussed in Sect.\ \ref{sec:4}. This procedure allows for all possible cancellations between contributions depending on different
matrix elements and gives the maximally conservative limit on the leptoquark couplings. Only with this strategy models can be reliably ruled
out based on the available information.
Given the large uncertainties in the  matrix elements, their precise treatment is consequential. While some of the theoretical
parameters can be argued to have a Gaussian distribution, for example the lattice values we use, and can hence be treated as Gaussian
nuisance parameters, this is certainly not true for others: specifically, several of the ranges above are obtained from the spread of
different available calculations. The idea here is simply to cover the full possible range for the parameter, but there
is no ``most likely'' value for it within this range. For these cases, we therefore assume these parameters to lie within the specified
range where they do not contribute to the $\chi^2$ and minimize the total $\chi^2$ under that assumption. 
This procedure is called Range-fit (R-fit) and was introduced in Ref.~\cite{Hocker:2001xe}.
\end{itemize}
In case of hadronic and diamagnetic EDMs the two strategies can lead to very different constraints, and the true constraints are
expected to lie in between these two extremes. It should be stressed that relatively modest improvement on the theoretical precision of
the matrix elements would essentially align the constraints obtained with the two strategies. Ref.~\cite{Chien:2015xha} showed that
theoretical control at the $50\%$ level would cause the ``Central" and ``R-fit" constraints to agree within a factor of two to three.

It turns out that in many cases one particular EDM measurement dominates the constraint. In order to illustrate which EDMs are
sensitive to which leptoquark interactions, we first give constraints for the individual EDM measurements assuming that a single
CP-violating source dominates at the high-energy scale. In order to make the resulting limits more transparent, in
Table~\ref{tab::hadsources} the dominant source on the hadronic level for each coupling is given. We will later discuss more global
scenarios.

\begin{table}
\centering{
\begin{tabular}{c|ccc}
$(X/Y)_{ab}$    & u              & c        & t\\\hline
$e$             & $C_{S,P,(T)}$      & $d_e$    & $d_e$\\[1.3ex]
$\mu$           & $d_{n,(p)}^{(a)},\bar g_{0,1}$ & $d_{n,(p)}^{(b)}$    & $d_\mu$\\[1.3ex]
$\tau$          & $d_{n,(p)}^{(a)},\bar g_{0,1}$ & $d_{n,(p)}^{(b)}$    & $d_\tau,d_{n,(p)}^{(c)}$
\end{tabular}
\caption{\label{tab::hadsources} Dominant contributions at the hadronic level for each combination of lepton and quark in the
$R_2$ and semileptonic $S_1$ scenarios. Note that all of the presently available observable classes give relevant constraints for at
least one possible coupling. $d_n^{(a)}$ denotes a contribution from the neutron EDM that stems from $d_u^{\rm eff}$ (and partly
$\tilde d_u$), $d_n^{(b)}$ denotes a contribution via charm quark EDM (or the Weinberg operator if the corresponding coefficient
should turn out to be much smaller than its present central value) and $d_n^{(c)}$ denotes a contribution from the Weinberg operator.
$\bar g_{0,1}$ are completely dominated by $\tilde d_u$ where they are relevant. Brackets indicate a contribution generically smaller than the
dominant one(s), but only within a factor of 10. Where several entries are listed, their hierarchy might depend on the values of
the corresponding hadronic matrix elements.} } 
\end{table}

\subsection{Constraints on individual leptoquark interactions}
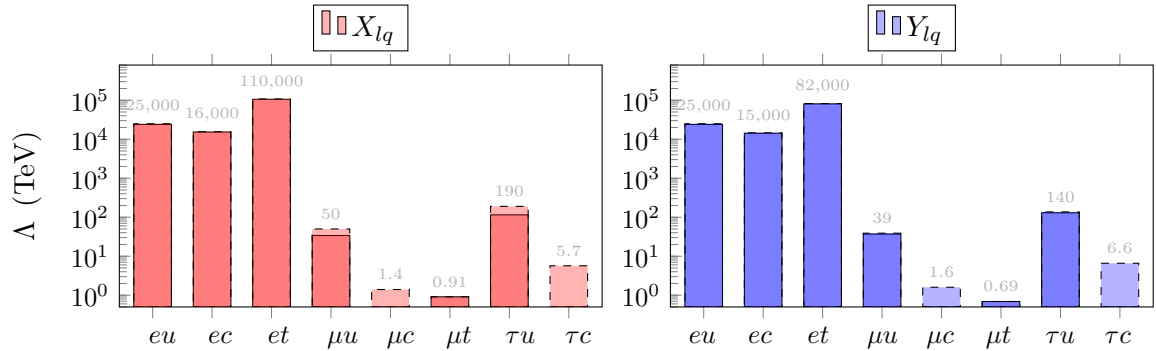
\begin{figure}[t!]\center
\pgfplotstableread[row sep=\\,col sep=&]{
    interval &  rfit&X   \\
    eu   & 24000&25000  \\
    ec   & 15300&15600 \\
    et    &106000&108000   \\
    muu & 34&50 \\
	muc & 0.12&1.4  \\
	mut &0.91& 0.91\\
	tauu & 115 & 190\\	
	asd & 0.001 &5.7  \\
	taut & 0.017 & 0.066\\
}\mydata
\begin{tikzpicture}
    \begin{axis}[
            ybar,ymode=log,
           log origin=infty    ,         
            bar width=.5cm, 
                        bar shift=0pt,
            width=.5\textwidth,
            height=.3\textwidth,
            legend style={at={(0.5,1.25)},
                anchor=north,legend columns=-1},
           symbolic  x coords={eu,ec,et,muu,muc,mut,tauu,asd,taut},
           enlarge x limits=.08,
            xtick=data,
            xticklabel style={align=center,text width=.1cm, font=\small},
            xticklabels={\parbox[r][.2cm][b]{.4cm}{$eu$},\parbox[r][.2cm][b]{.4cm}{$ec$},\parbox[r][.2cm][b]{.4cm}{$et$},\parbox[r][.2cm][b]{.4cm}{$\mu u$}, \parbox[r][.2cm][b]{.4cm}{$\mu c$}, \parbox[r][.2cm][b]{.4cm}{$\mu t$},\parbox[r][.2cm][b]{.4cm}{$\tau u$},\parbox[r][.2cm][b]{.4cm}{$\tau c$},\parbox[r][.2cm][b]{.4cm}{$\tau t$}},
            tick label style={font=\small},
            ytick={0.1,1,10,100,1000,10000,100000,1000000},
            ylabel={$\Lambda\,$ (TeV)}, 
         nodes near coords/.append style={font=\tiny},                    
              ymin=.5,ymax=800000,
                          ytick pos=left,
            nodes near coords align={vertical},            
            point meta=rawy  ]			
         \addplot[fill=red, fill opacity=.3]  table[x=interval,y=rfit]{\mydata};
        \addplot[nodes near coords={\pgfmathprintnumber[fixed,fixed zerofill,fixed relative, precision=2]{\pgfplotspointmeta}}, dashed, fill=red, fill opacity=.3]     table[x=interval,y=X]{\mydata};
       \legend{$X_{lq}$}
    \end{axis}

\end{tikzpicture}
\pgfplotstableread[row sep=\\,col sep=&]{
    interval &  rfitY&Y   \\
    eu   & 24000&25000  \\
    ec   & 14300&14600 \\
    et    &80800&82000   \\
    muu & 37&39 \\
    muc & 0.12&1.6\\
        mut &0.69& 0.69 \\
	tauu &130&137 \\
	tauc &0.00077& 6.6  \\
	taut & 0.017&0.066\\
    }\mydata
\begin{tikzpicture}
    \begin{axis}[
            ybar,ymode=log,
           log origin=infty    ,         
            bar width=.5cm, 
                        bar shift=0pt,
            width=.5\textwidth,
            height=.3\textwidth,
            legend style={at={(0.5,1.25)},
                anchor=north,legend columns=-1},
           symbolic  x coords={eu,ec,et,muu,muc,mut,tauu,tauc,taut},
           enlarge x limits=.08,
            xtick=data,
            xticklabel style={align=center,text width=1cm, font=\small},
            xticklabels={\parbox[r][.2cm][b]{.4cm}{$eu$},\parbox[r][.2cm][b]{.4cm}{$ec$},\parbox[r][.2cm][b]{.4cm}{$et$},\parbox[r][.2cm][b]{.4cm}{$\mu u$}, \parbox[r][.2cm][b]{.4cm}{$\mu c$}, \parbox[r][.2cm][b]{.4cm}{$\mu t$},\parbox[r][.2cm][b]{.4cm}{$\tau u$},\parbox[r][.2cm][b]{.4cm}{$\tau c$},\parbox[r][.2cm][b]{.4cm}{$\tau t$}},
            tick label style={font=\small},
            ytick={0.1,1,10,100,1000,10000,100000,1000000}, 
         nodes near coords/.append style={font=\tiny},                    
              ymin=0.5,ymax=800000,
                          ytick pos=left,
            nodes near coords align={vertical},            
            point meta=rawy  ]			
         \addplot[fill=blue, fill opacity=.3]  table[x=interval,y=rfitY]{\mydata};
        \addplot[nodes near coords={\pgfmathprintnumber[fixed,fixed zerofill,fixed relative, precision=2]{\pgfplotspointmeta}}, dashed, fill=blue, fill opacity=.3]     table[x=interval,y=Y]{\mydata};
       \legend{$Y_{lq}$}
    \end{axis}

\end{tikzpicture}
\caption{\small The figure summarizes the most stringent constraints on the scale of $R_2$ and $S_1$ leptoquarks,  assuming
$X_{lq}=1/\Lambda^2$ and $Y_{lq}=1/\Lambda^2$, in the left and right panels, respectively. The dashed bars show the constraints
obtained using the `central' strategy while the solid bars indicate those obtained in the `R-fit' approach. The limit on the scale for
the $X_{\tau t}$ and $Y_{\tau t}$ couplings are weaker than the range of $\Lambda$ shown in the panels.}
\label{fig:XY_chart}
\end{figure}

\begin{figure}[t!]\center
\pgfplotstableread[row sep=\\,col sep=&]{
    interval &  rfit&X   \\
    eu   & 0.005&58  \\
    ec   & .001&195\\
    et    &427&1750   \\
    muu & 275&1090 \\
    muc & 4&15  \\
        mut &0.001& 32  \\
tauu &2090&8100 \\
tauc &45& 103  \\
taut & 14&43\\
    }\mydata
\begin{tikzpicture}
    \begin{axis}[
            ybar,ymode=log,
           log origin=infty    ,         
            bar width=.5cm, 
                        bar shift=0pt,
            width=.6\textwidth,
            height=.33\textwidth,
            legend style={at={(0.5,1.25)},
                anchor=north,legend columns=-1},
           symbolic  x coords={eu,ec,et,muu,muc,mut,tauu,tauc,taut},
           enlarge x limits=.08,
            xtick=data,
            xticklabel style={align=center,text width=1cm, font=\small},
            xticklabels={\parbox[r][.2cm][b]{.4cm}{$ud$},\parbox[r][.2cm][b]{.4cm}{$us$},\parbox[r][.2cm][b]{.4cm}{$ub$},\parbox[r][.2cm][b]{.4cm}{$cd$}, \parbox[r][.2cm][b]{.4cm}{$cs$}, \parbox[r][.2cm][b]{.4cm}{$cb$},\parbox[r][.2cm][b]{.4cm}{$td$},\parbox[r][.2cm][b]{.4cm}{$ts$},\parbox[r][.2cm][b]{.4cm}{$tb$}},
            tick label style={font=\small},
            ytick={0.1,1,10,100,1000,10000},
            ylabel={$\Lambda\,$ (TeV)}, 
         nodes near coords/.append style={font=\tiny},                    
              ymin=5,ymax=30000,
                          ytick pos=left,
            nodes near coords align={vertical},            
            point meta=rawy  ]			
         \addplot[fill=blue, fill opacity=.3]  table[x=interval,y=rfit]{\mydata};
        \addplot[nodes near coords={\pgfmathprintnumber[fixed,fixed zerofill,fixed relative, precision=2]{\pgfplotspointmeta}}, dashed, fill=blue, fill opacity=.3]     table[x=interval,y=X]{\mydata};
       \legend{$Z_{ud}$}
    \end{axis}

\end{tikzpicture}
\caption{\small The figure summarizes the most stringent constraints on the scale of the $S_1$ leptoquark, assuming
$Z_{ud}=1/\Lambda^2$. The dashed bars show the constraints obtained using the `central' strategy while the solid lines indicate those
obtained in the `R-fit' approach.}
\label{fig:Z_chart}
\end{figure}
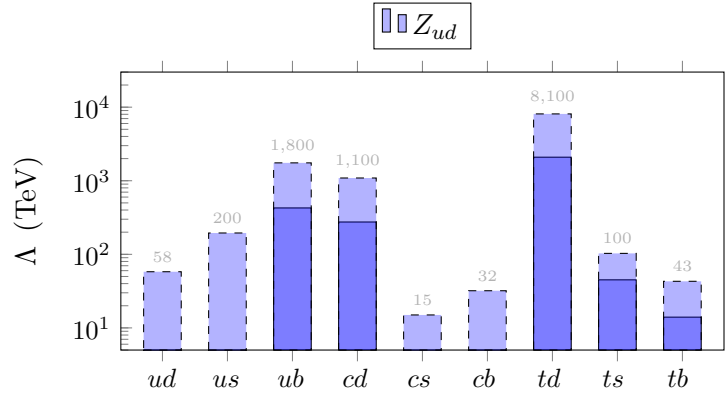

The limits on the combinations $X_{lq}$ (limits on $Y_{lq}$ are similar and therefore not shown) are collected in Tables
\ref{tab:CentralLimitsR2} and \ref{tab:RfitLimitR2} for the Central and R-fit strategies, respectively. Limits on $Z_{qq'} $ are shown
in Tables \ref{tab:CentralLimitS1Quark} and \ref{tab:RfitLimitS1Quark}. Here, we have assumed $m_{LQ}=1$ TeV, and give constraints on
the dimensionless combinations $\bar{X}_{lq}\equiv m^2_{LQ} X_{lq}$ and $\bar{Z}_{qq'}\equiv m^2_{LQ} Z_{qq'}$. Increasing or
decreasing $m^2_{LQ}$ will roughly decrease or increase the limits by the same amount modulo $\mathcal O(1)$ RGE factors due to the
slightly different evolution from $m^2_{LQ}$ to the electroweak scale. In principle, we can turn the strategy around and set a lower
bound on the mass of $m_{LQ}$ by assuming that the dimensionless couplings $\bar{X}_{lq}$ and $\bar{Z}_{qq'}$ are numbers of $\mathcal
O(1)$. We show the limits on the leptoquark scale obtained in this way in Figs.\ \ref{fig:XY_chart} and \ref{fig:Z_chart}. We stress,
however, that these figures mainly serve as a way to visualize the constraints, as the limits on $\Lambda$ cannot generally be
interpreted as limits on $m_{LQ}$. In fact, naturalness considerations suggest that the dimensionless couplings, $\bar X_{lq}$, $\bar
Y_{lq}$, and $\bar Z_{ud}$, can be very small \cite{Arnold:2013cva}. Similar small dimensionless couplings appear in models where a
version of minimal flavor violation is assumed \cite{Davidson:2010uu}. 

In all tables we have removed limits that are much larger than $\{\bar X_{lq} ,\,\bar Z_{qq'}\} > (4\pi)^2 $. Such bounds cannot be trusted since  they would indicate large (non-perturbative) dimensionless couplings or $m_{LQ}\ll 1$ TeV (at which point the EFT would break down). In some cases, we nevertheless provide naive limits to see by how much the EDM must improve to set relevant constraints.  
We also provide constraints from potential future proton, deuteron, and radium EDM measurements using the sensitivities in
Table~\ref{tab:exptfut}. The impact of improved limits from the other EDMs can be obtained by rescaling the entries in the Tables.

\begin{table}
\footnotesize\center
$
\begin{array}{c|cccccccccccc}
{\rm Cent.}& \bar X_{{e}u} & \bar X_{{e}c} & \bar X_{{e}t}& \bar X_{\mu u} & \bar X_{\mu c}& \bar X_{\mu t}& \bar X_{\tau u} & \bar X_{\tau c}& \bar X_{\tau t} \\\hline
 d_{\mu } & - & - & - & - & 60 & 1 & - & - & - \\
 d_{\tau } & - & - & - & - & - & - & - & - & 300 \\
 d_n & 0.1 & 200\,(-) & - & 8 \cdot 10^{-4} &1 \,(6)& - & 7 \cdot 10^{-5} & 7\cdot 10^{-2}\,(0.3) & 
400 \\
 d_{\text{Hg}} & 1\cdot 10^{-8} & 2\cdot 10^{-7} & 4\cdot 
10^{-9} & 4 \cdot 10^{-4} & 0.5\,(3) & - & 3 \cdot 10^{-5} & 3 \cdot 10^{-2}\,(0.2) &200 \\
 d_{\text{Tl}} & 3\cdot 10^{-7} & 6\cdot 10^{-7} & 1\cdot 
10^{-8} & - & - & - & - & - & - \\
 \text{YbF} & 3\cdot 10^{-7} & 5\cdot 10^{-7} & 1\cdot 10^{-8} \
& - & - & - & - & - & - \\
 \text{HfF} & 3\cdot 10^{-8} & 5\cdot 10^{-8} & 1\cdot 10^{-9} \
& - & - & - & - & - & - \\
 \text{ThO} & 2\cdot 10^{-9} & 4\cdot 10^{-9 }& \ 9\cdot 10^{-11} 
& - & - & - & - & - & - \\
\hline
 d_{p,\, {\rm fut}} & 8 \cdot 10^{-6} & 5 \cdot 10^{-2}\,(0.4) & 500 & 6\cdot 10^{-8} & 3\cdot 10^{-4} \,(2\cdot 10^{-3})& 2 & 5\cdot 10^{-9} & 2 \cdot 10^{-5} \,(1 \cdot 10^{-4})& 0.1 \\
 d_{D,\, {\rm fut}} & 1 \cdot 10^{-5}& 3 \cdot 10^{-2} \,(90)& - & 1\cdot 10^{-7} & 2\cdot 10^{-4} \,(0.4)& 30 & 1\cdot \
10^{-8} &1 \cdot 10^{-5}\,(3\cdot 10^{-2})& 2 \\
 d_{\text{Ra},\, {\rm fut}} & 4\cdot 10^{-2} & - & - &2 \cdot 10^{-4} & 200 & - & 1\
\cdot 10^{-5} & 10& 800 \\
\end{array}
$
\caption{Limits on the $R_2$ couplings  $\bar X_{{l}q} \equiv m_{LQ}^2 X_{{l}q}$ from different EDM measurements. We took $m_{LQ}=1$
TeV and assumed the central values  for all matrix elements. Given the uncertain nature of the charm tensor charge we also show the
limits obtained with $g^c_T\to 0$ in brackets, whenever this has an impact. The last three rows show the expected limits from future
experiments. }
\label{tab:CentralLimitsR2}
\end{table}

\begin{table}
\small\center
$
\begin{array}{c|ccccccccc}
{\text{ R-fit}}& \bar X_{{e}u} & \bar X_{{e}c} & \bar X_{{e}t}& \bar X_{\mu u} & \bar X_{\mu c}& \bar X_{\mu t}& \bar X_{\tau u} & \bar X_{\tau c}& \bar X_{\tau t} \\\hline
 d_{\mu } & - & - & - & - & 60 & 1 & - & - & - \\
 d_{\tau } & - & - & - & - & - & - & - & - & 300 \\
 d_n & 0.1 & - & - & 9\cdot 10^{-4} & - & - & 7\cdot 10^{-5} & 
- & - \\
 d_{\text{Hg}} & 1\cdot 10^{-8} & - & - & - & - & - & - & - & - \\
 d_{\text{Tl}} & 3\cdot 10^{-7} & 6\cdot 10^{-7} & 1\cdot \
10^{-8} & - & - & - & - & - & - \\
 \text{YbF} & 4\cdot 10^{-7} & 5\cdot 10^{-7} & 1\cdot 10^{-8} \
& - & - & - & - & - & - \\
 \text{HfF} & 3\cdot 10^{-8} & 5\cdot 10^{-8} & 1\cdot 10^{-9} \
& - & - & - & - & - & - \\
 \text{ThO} & 2\cdot 10^{-9} & 4\cdot 10^{-9} & 9\cdot 10^{-11} \ 
& - & - & - & - & - & - \\
\hline
 d_{p,\,{\rm fut}} & 9\cdot 10^{-6} & - & - & 7\cdot 10^{-8} &-& 20 & 6\cdot 10^{-9} &-& 1 \\
 d_{D,\,{\rm fut}} & 2 \cdot 10^{-5} & - & - & 4 \cdot 10^{-7} & - & - & - & - & - \\
 d_{\text{Ra},\,{\rm fut}} & 0.5 & - & - & 2\cdot 10^{-3} & - & \
- & 1\cdot 10^{-4} & 300 & - \\
\end{array}
$
\caption{Limits on the $R_2$ couplings $\bar X_{{l}q} \equiv m_{LQ}^2 X_{{l}q}$ from different EDM measurements. We took $m_{LQ}=1$ TeV and varied the matrix elements within their allowed ranges to get conservative constraints.  }
\label{tab:RfitLimitR2}
\end{table}
We start with by discussing constraints on the $\bar{X}_{lq}$ combinations. For the couplings involving electrons, $\bar{X}_{eq}$, the
relevant low-energy operators are the electron EDM and CP-odd electron-nucleon interactions, only. These are probed efficiently by the
paramagnetic systems and, because of its impressive experimental limit, 
also by the Hg EDM. For the $\bar{X}_{eu}$ couplings the electron-nucleon interactions are induced at tree-level and are dominant.
As a result, the ThO experiment limits $\bar{X}_{eu}$ at the $10^{-9}$ level in agreement with the analysis of
Ref.~\cite{Fuyuto:2018scm}.
The Hg constraint is not too far from the ThO one; interestingly the limit stems from the combination of $C_{S}$ and $C_P$ while
the contribution from $C_T$ is about an order of magnitude smaller. This result is surprising at first, since the atomic
coefficient of $C_T$ is two orders of magnitude larger than the one of $C_S$ and $C_P$ is in many cases completely neglected. This goes
to show again that only the combination of the various hierarchies allows to judge the relevance of a given contribution.

For $\bar X_{ec}$ and $\bar X_{et}$ the limits 
are dominated by the one-loop contributions to the electron EDM, which receives a relative $m_{c,t}/m_e$ enhancement compared to
the other loop contributions,
leading to constraints at the $10^{-9,-10}$ level for these couplings as well. For the Central strategy, the limits from Hg are
only a factor of five weaker.

Moving to the R-fit limits, we see that for the paramagnetic systems the limits on $\bar{X}_{eq}$ are barely affected. This is not
surprising as the theoretical control over the atomic matrix elements is very good. Experimental progress in paramagnetic
systems will therefore directly translate into stronger bounds on the respective couplings. The Hg constraints are significantly
affected for the $\bar{X}_{ec}$ and $\bar{X}_{et}$ couplings, because the contribution from the electron EDM is poorly understood, see
Eq.~\eqref{dHg}. Work is in progress to improve the associated atomic theory \cite{MartinInprogress}. Once this is achieved,
progress in Hg will improve the bounds for all three couplings as well for both strategies.

Turning to the muonic couplings, $\bar{X}_{\mu q}$, the picture changes drastically. The paramagnetic systems play no role as no
significant contributions to the electron EDM or electron-nucleon couplings are induced. On the other hand the muon mass is still
too small to induce large hadronic couplings. For the $\bar X_{\mu u}$ coupling sizable contributions to up-quark EDM and
chromo-EDM are generated and these dominate the neutron and Hg EDMs. The resulting limits are at the $10^{-3,-4}$ level. 

For the $\bar{X}_{\mu c}$ the situation is rather complicated: a sizable muon EDM is induced, but since its experimental limit is
relatively weak, so is the resulting constraint on the coupling.  There are two-loop contributions to the Weinberg operator that are
suppressed by the muon mass but still contribute sizably to the neutron and Hg EDMs. The resulting limits are at the $\mathcal O(1)$
level and given in brackets. Finally, there is a sizable contribution to the charm EDM, contributing to the nucleon EDMs,
which we treat as described above, yielding a stronger constraint for the Central strategy.

In case of $\bar{X}_{\mu t}$, the Weinberg contributions are suppressed by $1/m_t$ and negligible at present,
while at the same time the contribution to the muon EDM gets enhanced due to the large top mass. The current muon EDM limit then
constrains $\bar{X}_{\mu t} \lesssim \mathcal{O}(1)$. 

The uncertainties for Hg are large enough to completely remove the constraints in the R-fit approach, showing the importance of
improved calculations. For the up coupling, the limit from the neutron EDM is only slightly weakened.
The Weinberg contribution from $\bar X_{\mu c}$ worsens by a factor of five, while the contribution via the charm EDM is allowed to
vanish, but can at the same time  cancel the Weinberg contribution, leaving no limit in the R-fit case.  As the muon EDM limit is not affected by theoretical uncertainties, its constraints do not change in the
R-fit approach.

The limits on $X_{\mu {u,c}}$ will be improved by future experiments on hadronic systems; for instance $d_p$ and/or $d_D$ can
potentially improve them by several orders of magnitude. These experimental developments should be matched by improved
determinations of the corresponding matrix elements, especially for the charm EDM contribution. For $X_{\mu t}$, the strength of
the muon EDM will not be matched even by $d_{p,D}$; hadronic uncertainties are not an issue here, so experimental progress will
immediately translate into improved knowledge of this coupling.

For the couplings to the $\tau$ the pattern is similar to the muon case. $\bar{X}_{\tau u }$ leads to large up-quark
\mbox{(chromo-)EDMs}, while $\bar{X}_{\tau c}$ and $\bar{X}_{\tau t}$ both induce the Weinberg operator, together with the charm
EDM in the former and the  $\tau$ EDM in the latter case. The best limits in the Central approach presently come from the Hg and
neutron EDM, with the $\tau$ EDM being competitive for the coupling to the top, but all existing limits are rather weak. 

In the R-fit approach the limit on $X_{\tau u}$ from the neutron EDM is mostly unaffected, while cancellations become possible for $X_{\tau
c,t}$, again highlighting the importance of improved matrix element determinations. The  $\tau$ EDM presently remains as the only and hence best limit on$X_{\tau t}$, although the constraint is too weak to be of significance.

Importantly, the $\tau$ EDM could be improved already with existing data, since the existing bound stems from only $\sim
30\,{\rm fm}^{-1}$ of Belle data. Belle~II  could then improve the $\tau$-EDM by one to two orders of magnitude
\cite{Kou:2018nap,Chen:2018cxt} which would provide a constraint at the $\mathcal O(1)$ level. The $X_{\tau u,c}$ couplings can
be improved with future experiments on hadronic systems, with future storage-ring experiments on $d_p$ and $d_D$ potentially providing
constraints at the percent level.

To summarize our results for semileptonic leptoquark couplings, the constraints on the electron couplings are, not surprisingly, by far the
strongest and are dominated by paramagnetic systems. For the couplings to heavier leptons, the up-quark interactions are well constrained by the neutron EDM both in the Central
and R-fit approach. The couplings to the charm are still reasonably well constrained in the Central strategy, but the hadronic and
nuclear uncertainties are significant, as can be seen from R-fit limits. For the couplings $\bar{X}_{\mu t}$ and $\bar{X}_{\tau t}$
the current muon and $\tau$ EDM limits are not strong enough yet to set significant constraints, but this is expected to change
in the near future. We note that the  Hg EDM would provide a great all-in-one system if hadronic, nuclear, and atomic
theory could be improved, as it provides strong Central constraints on almost all leptoquark couplings. It is interesting that the
$\bar{X}_{lq}$ interactions provide a rich enough structure that essentially all different classes  of EDM experiments play a role. 
Ongoing experimental efforts aim at improving the sensitivity by at least one order of magnitude for all couplings involved;
notably, achieving the challenging goals of the proton- and deuteron-EDM experiments would improve the sensitivity for some of the
couplings involving heavier leptons by several orders of magnitude.

\begin{table}[t]
\small
$
\begin{array}{c|ccccccccc}
 {\rm Cent.} & \bar Z_{ud} & \bar Z_{us} & \bar Z_{ub} &\bar Z_{cd} & \bar Z_{cs} &\bar Z_{cb} & \bar Z_{td} & \bar Z_{ts} &\bar Z_{tb} \\\hline
 d_n & 3\cdot 10^{-4} & 9\cdot 10^{-5} & 3\cdot 10^{-6} & 5\cdot 10^{-6} & 9\cdot 10^{-3} \
& 2\cdot 10^{-3} & 1\cdot 10^{-7} & 2\cdot 10^{-4} & 9 \cdot 10^{-4} \\
 d_{\text{Hg}} & 4\cdot 10^{-4} & 3\cdot 10^{-5} & 3\cdot 10^{-7} & 8\cdot 
10^{-7} & 4\cdot 10^{-3} & 1\cdot 10^{-3} & 2\cdot 10^{-8} & 9\cdot 10^{-5} & 5\cdot 10^{-4}\\
\hline
d_{p,\,{\rm fut}} & 9\cdot 10^{-8} & 1\cdot 10^{-8} & 4\cdot 10^{-10} & 3\cdot 10^{-9} & 4\cdot 10^{-6} & 6\cdot 10^{-7} & 5\cdot \
10^{-11} & 7\cdot 10^{-8} & 3\cdot 10^{-7} \\
 d_{D,\,{\rm fut}} & 2\cdot 10^{-7} & 6\cdot 10^{-9} & 1\cdot 10^{-10} & 4\cdot 10^{-10} & 2\cdot 10^{-6} & 1 \cdot 10^{-5}\,(1 \cdot 10^{-4}) & 8\cdot 10^{-12} & 3\cdot 10^{-8} &1 \cdot 10^{-5}\\
 d_{\text{Ra},\,{\rm fut}}& 8\cdot 10^{-5} & 3\cdot 10^{-6} & 5\cdot 10^{-8} & 2\cdot 10^{-7} & 2 & 5\cdot 10^{-2} & 3\cdot 10^{-9} & -& 5\cdot 10^{-3} \\
\end{array}
$
\caption{Limits on the $S_1$ di-quark couplings $\bar Z_{qq'} \equiv m_{LQ}^2 Z_{qq'}$ for $m_{LQ}=1$ TeV using central values for all
matrix elements. The limits in brackets  denote constraints obtained with $g^c_T\to 0$.}
\label{tab:CentralLimitS1Quark}
\end{table}

\begin{table}[t]
\small\center
$
\begin{array}{c|ccccccccc}
 {\text{R-fit}}& \bar Z_{ud} & \bar Z_{us} & \bar Z_{ub} &\bar Z_{cd} & \bar Z_{cs} &\bar Z_{cb} & \bar Z_{td} & \bar Z_{ts} &\bar Z_{tb} \\\hline
 d_n & - &-
  & 5\cdot 10^{-6} & 1\cdot 10^{-5} & 6\cdot 10^{-2} &- & 2\cdot 10^{-7} & 5\cdot 10^{-4} & 5\cdot 10^{-3} \\
 d_{\text{Hg}} & - & - & - & - & -& - & - &- & - \\
 \hline
 d_{p,\,{\rm fut}} & - &-
 & 7\cdot 10^{-10} & 7\cdot 10^{-9} & \
2\cdot 10^{-5} & - & 1\cdot 10^{-10} & 2\cdot 10^{-7} & \
2\cdot 10^{-6} \\
 d_{D,\,{\rm fut}} & - & -
  & 4\cdot 10^{-10} & 1\cdot 10^{-9} & \
6\cdot 10^{-6} & - & 3\cdot 10^{-11} & 9\cdot 10^{-8} & - \\
 d_{\text{Ra},\,{\rm fut}} & - & -
 &8\cdot 10^{-7} & 1\cdot 10^{-6} & \
70 & 2 & 3\cdot 10^{-8} & - & 0.2 \\
\end{array}
$
\caption{Limits on the $S_1$ di-quark couplings  $\bar Z_{{q}q'} \equiv m_{LQ}^2 Z_{{q}q'}$ for $m_{LQ}=1$ TeV. We varied  the matrix elements within their allowed ranges. }
\label{tab:RfitLimitS1Quark}
\end{table}

Finally, we discuss the $\bar{Z}_{qq'}$ limits given in Tables~\ref{tab:CentralLimitS1Quark} and \ref{tab:RfitLimitS1Quark}.
As these couplings only induce hadronic EDMs, they are at present all dominated by
either $d_n$ or $d_{\mathrm{Hg}}$.
Couplings involving one light quark
induce large contributions to up and down chromo-EDMs and are dominated by $d_{\mathrm{Hg}}$, because of the pion-exchange
contributions to the atomic EDM. Similarly large contributions to $\bar g_{0,1}$ are expected from the four-quark operators induced by $Z_{us}$. However, this is effect is mitigated by the fact that $C_{quqd}^{(1)}\simeq C_{quqd}^{(8)}$ at $\mu =1$ GeV, see Table \ref{tab:RunningS1Quark},	leading to partial cancellations even in the Central approach. 
The other couplings mainly induce the Weinberg operator ($Z_{cb, tb}$), the strange (C)EDMs ($Z_{ts,cs}$), or the four-quark operator ($Z_{ud}$. 
None of these contributions generate an enhanced $\bar g_{0,1}$, so that the
bounds from $d_n$ and $d_{\mathrm{Hg}}$ on these couplings are comparable. 
In addition, $Z_{cd,cs,cb}$ induce the charm EDM, which has a far smaller impact than was the case for the $X_{lq}$ and $Y_{lq}$
couplings. The effect is only visible for the potential future constraint on $Z_{cb}$ from $d_D$, which is due to the fact that
contributions via  the Weinberg operator cancels in this system.

 In the R-fit approach, the limits soften significantly. For the $\bar{Z}_{ud,us}$ the bounds essentially disappear because the matrix
 element of the CP-odd four-quark operators are poorly understood.
All other couplings are still significantly constrained by $d_n$.

Future experiments with the neutron, light nuclei and diamagnetic systems can improve the limits significantly.
We note that in our projections $\bar{Z}_{cs}$, $\bar{Z}_{cb}$, $\bar{Z}_{ts}$ and $\bar{Z}_{tb}$ are barely limited by
$d_{\mathrm{Ra}}$. To a large extent this can be explained by our theoretical ignorance. These couplings mainly induce nucleon EDMs,
which do not appear in Eq.~\eqref{dRa} because their contributions to the Ra EDM are expected to be small with respect to the
pion-exchange terms. The  $d_{\mathrm{Ra}}$ limits in Tables~\ref{tab:CentralLimitS1Quark} and \ref{tab:RfitLimitS1Quark} are therefore
not reliable for  couplings where $\bar g_0$ and $\bar g_1$ are not induced.

\subsection{Interplay of couplings}\label{sec:interplay} 
The above analysis of EDM constraints was based on the assumption that a single combination of leptoquark interactions is dominant at
the high-energy scale. This scenario, while easy to analyze, is not very realistic. In fact, in most models of leptoquarks,
interactions among different quarks and leptons are generated and possibly related by a flavor symmetry. It is therefore interesting to
study the complementarity of different EDM measurements by studying scenarios in which multiple CP-odd couplings are generated. 
Ideally, we would perform a global analysis where all possible CP-odd combinations for each leptoquark representation are considered
simultaneously. Without additional input on CP violation from non-EDM measurements, such a fit would not lead to any constraints as
there are more couplings than independent EDM measurements. We therefore limit ourselves to more constrained scenarios where only a few
couplings are turned on simultaneously.
The discussion of a more specific model motivated by the $B$ anomalies is deferred to the next subsection. In the following, we
will consider central values of the matrix elements and thus neglect the associated uncertainties.

\begin{figure}
\centering{
\includegraphics[scale = .7]{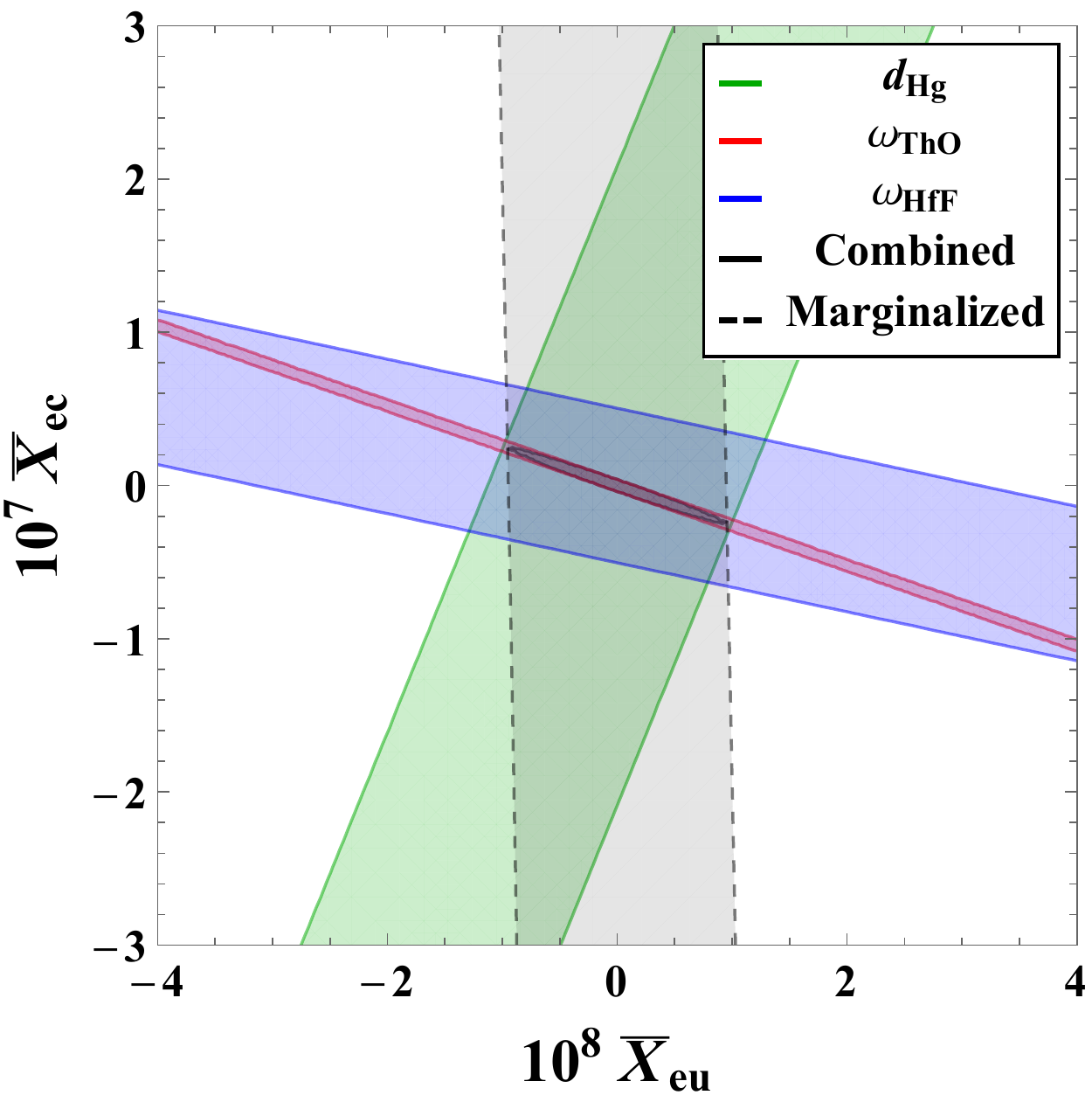}
}
\caption{$90\%$ C.L.\ constraints from various EDM experiments assuming only the $\bar X_{e u}$ and $\bar X_{e c}$ couplings are
present. The dark ellipse is the combined allowed region. The gray dashed band illustrates the allowed region when we also allow for a
nonzero $\bar X_{e t}$ coupling.}
\label{LQ12}
\end{figure}

We begin by analyzing scenarios involving electron couplings $\bar X_{e U}$ with $U=\{u,\,c,\,t\}$. Table \ref{tab:CentralLimitsR2}
shows that constraints on these couplings are individually dominated by the ThO measurement, while limits from HfF and Hg are slightly
weaker. Fig.~\ref{LQ12} shows the region in the $\bar X_{e u}$-$\bar X_{ec}$ plane that is allowed by the different EDM experiments.
The ThO and HfF constraints illustrate the fact that paramagnetic systems constrain similar combinations of $d_e\sim X_{ec,t}$ and
$C_S\sim \bar X_{eu}$, see \cite{PhysRevA.84.052108,Jung:2013mg,Fleig:2018bsf,Chupp:2014gka} for detailed discussions. As such, the allowed regions
for these two experiments overlap to a large extent. On the other hand, $d_\text{Hg}$ is sensitive to a different combination of $d_e$
and $C_S$, and thus provides a complimentary constraint \cite{Jung:2013mg,Fleig:2018bsf}, leading to stringent limits on both $\bar
X_{e u}$ and $\bar X_{e c}$.

Since the semileptonic CP-odd operators are all dominated by $\bar X_{eu}$, only $d_e$ is available to constrain both $\bar
 X_{ec,t}$. It is therefore not possible to obtain constraints on these two couplings individually. The linear combination that is
 constrained is in principle $\bar X_{ec}+m_t/m_c \bar X_{et}$, however, this is changed by renormalization group effects to $\bar
 X_{eQ}\approx \bar X_{ec}+4.4m_t/m_c \bar X_{et}$ at $\sim 1$~GeV. In the presence of all three coefficients, the plot remains
 identical with the replacement $\bar X_{ec}\to \bar X_{eQ}$.  
We see no immediate way for future EDM experiments to break this $\bar X_{e c}$-$\bar X_{e t}$ degeneracy.

\begin{figure}
\centering{
\includegraphics[scale = .55]{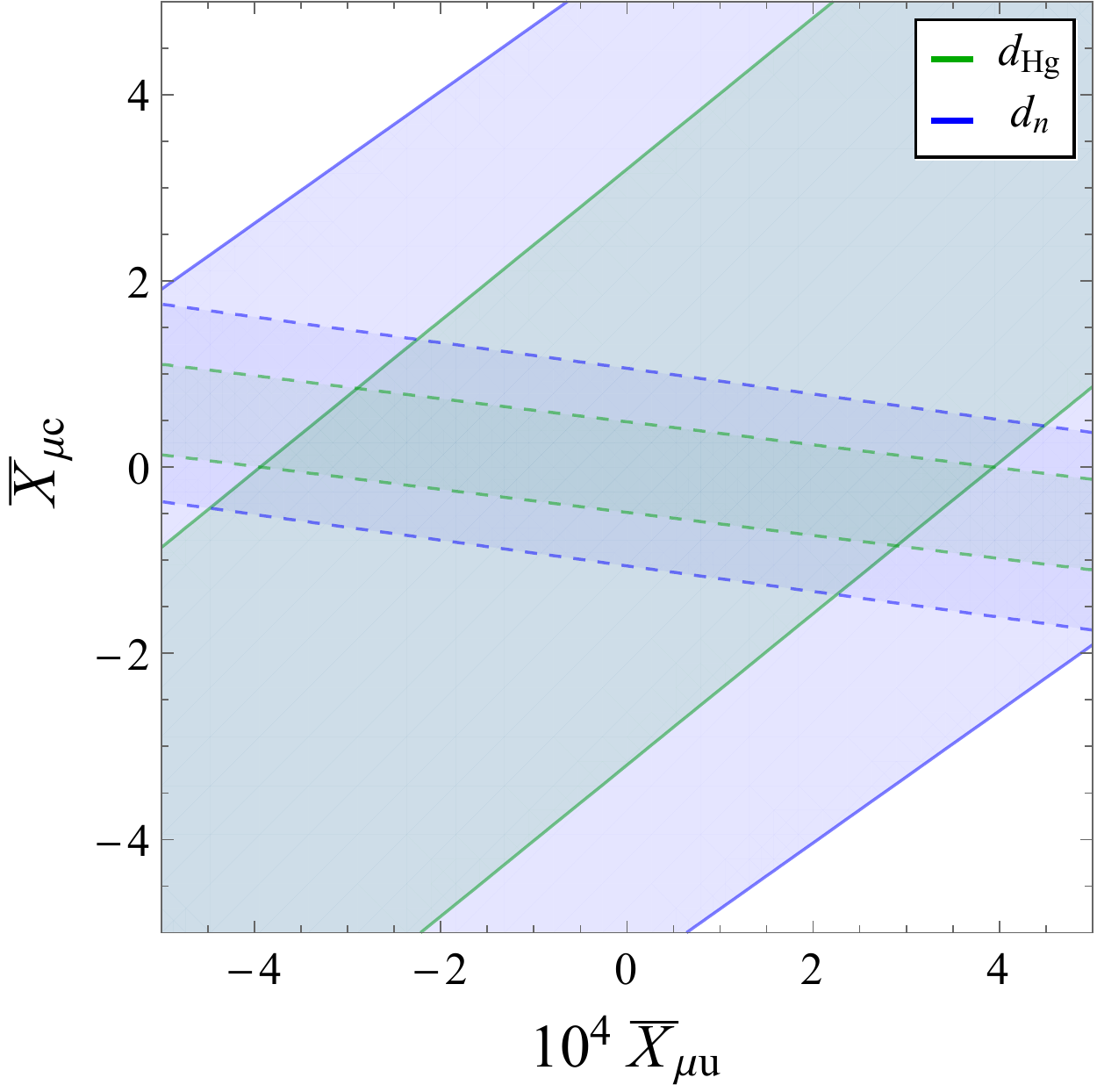}\hfill
\includegraphics[scale = .55]{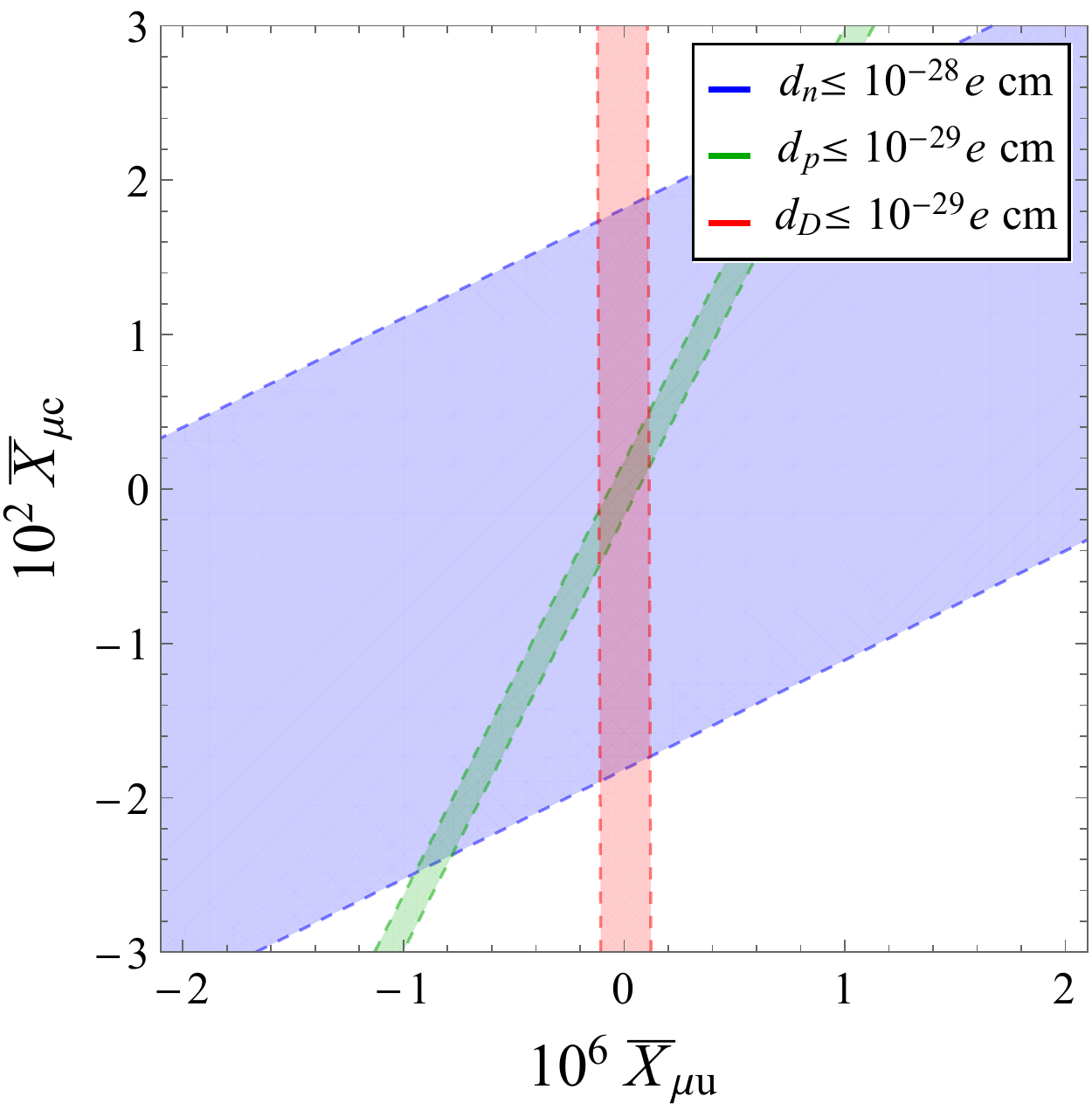}
}
\caption{Left panel: Current constraints from various EDM experiments assuming only the $\bar X_{\mu u}$ and $\bar X_{\mu c}$ couplings
are present. Solid (dashed) lines indicate the constraints for $g_T^c=0$ ($g_T^c= \frac{m_s}{m_c} g_T^s$). Right panel: constraints using the expected sensitivities of prospected
neutron, proton, and deuteron EDM experiments using  $g_T^c=0$. }
\label{LQ45}
\end{figure}

We can perform a similar analysis for the muonic couplings $\bar X_{\mu  U}$ with $U=\{u,\,c,\,t\}$. The couplings $\bar X_{\mu u}$ and
$\bar X_{\mu c}$ are mainly constrained by the neutron and Hg EDMs. These EDMs depend on very similar linear combinations of $\bar
X_{\mu u}$ and $\bar X_{\mu c}$, owing to the fact that the nucleon EDMs enter the Mercury constraint via the Schiff moment. This
degeneracy is in principle broken by the contribution from $\bar g_1$ in Hg, but only weakly. Consequently, an approximate free
direction emerges as depicted in the left panel of Fig.~\ref{LQ45}, showing the constraints from $d_n$ and $d_{\rm Hg}$ both for
 $g_T^c= \frac{m_s}{m_c} g_T^s$ and $g_T^c=0$.
This approximate degeneracy could be resolved with future experiments involving protons and/or deuterons  which would improve the
current limits by several orders of magnitude, and could distinguish between the two couplings, as can be seen in the right panel of
Fig.~\ref{LQ45}. Note that the dependence of $d_D$ on only $\bar X_{\mu u}$ apparent in this figure is an artefact of using only
the central values for the Weinberg matrix elements, which are equal in magnitude and have opposite signs. 
This cancellation does not take place when taking
a nonzero value for the charm tensor charge into account. This would also affect the slopes of the $d_{n,p}$ exclusion bands in the right panel; however,
the complementarity of the different systems remains intact in this scenario.

\begin{figure}
\centering{
\includegraphics[scale = .56]{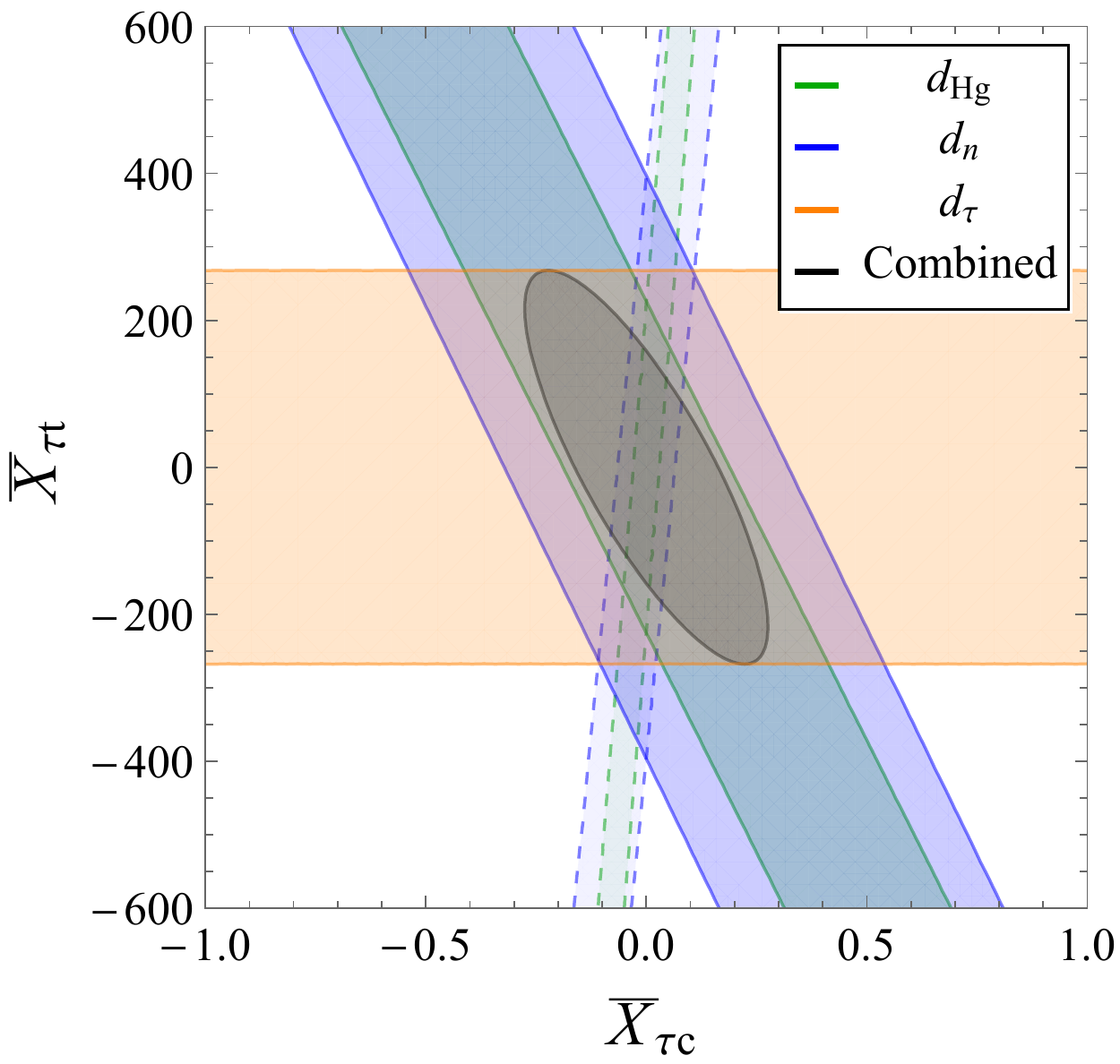}\hfill
\includegraphics[scale = .55]{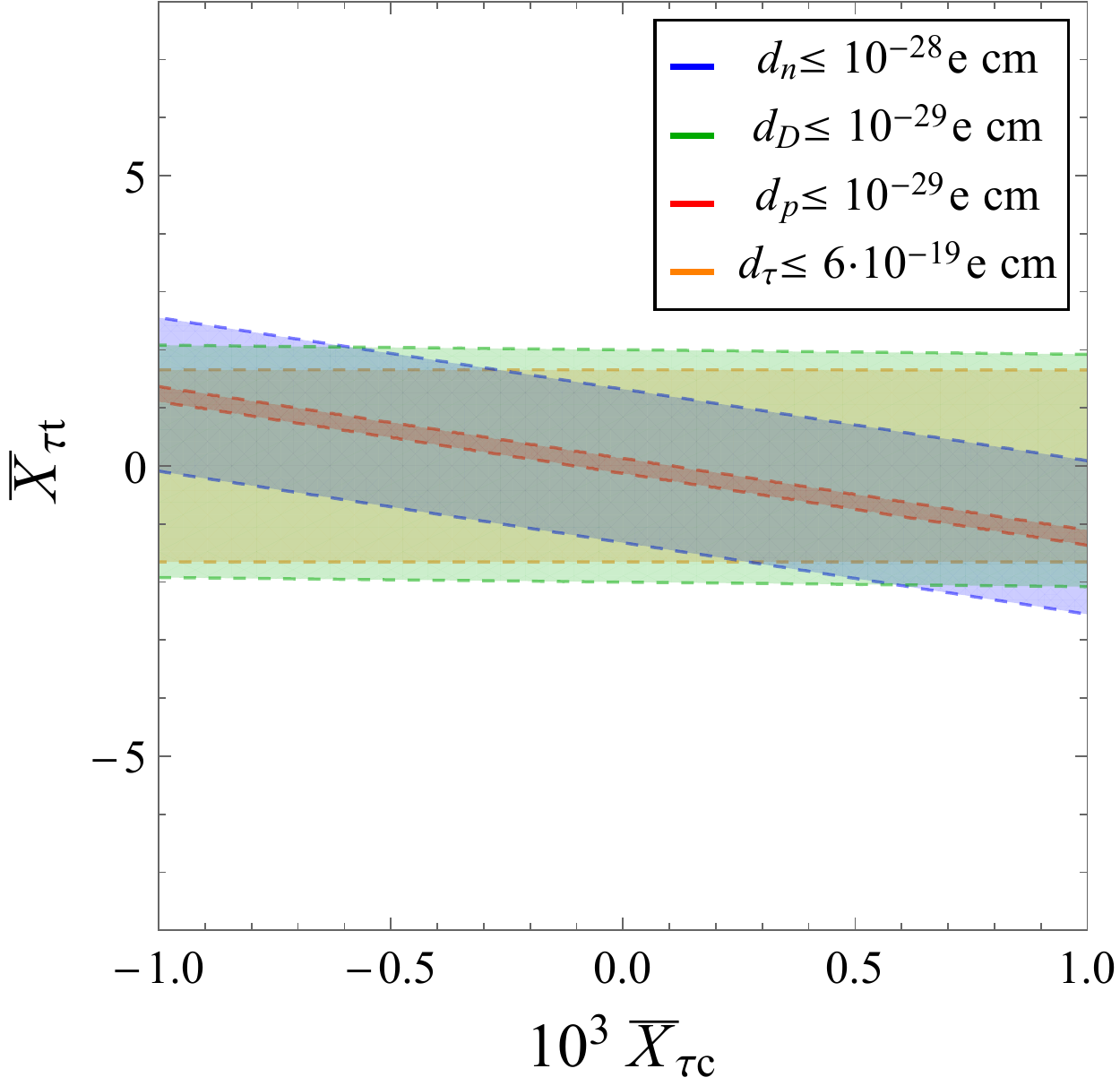}
}
\caption{Left panel: Constraints from various EDM experiments in the $\bar X_{\tau c}$-$\bar X_{\tau t}$ plane. The dark ellipse is the
combined allowed region. Solid (dashed) lines indicate the constraints for $g_T^c=0$ ($g_T^c= \frac{m_s}{m_c} g_T^s$). Right
panel: constraints using expected sensitivities on neutron, proton, deuteron, and $\tau$ EDM experiments using  $g_T^c=0$.}
\label{LQ78}
\end{figure}

For the tau couplings, $\bar X_{\tau U}$ with $U=\{u,\,c,\,t\}$, the $\bar X_{\tau u}$-$\bar X_{\tau c}$ plots look very similar to the
$\bar X_{\mu u}$-$\bar X_{\mu c}$ plots in Fig.~\ref{LQ45}. We therefore show contours in the $\bar X_{\tau c}$-$\bar
X_{\tau t}$ plane in Fig.~\ref{LQ78}. The neutron and Hg EDMs allow for a free direction for the same reason as above, which, in
principle, is removed by the current limit on the $\tau$ EDM. However, this still allows for large $\mathcal O(10^2)$ values of $\bar
X_{\tau t}$, and our analysis is not reliable for such large couplings. Future improvements would remedy this situation as shown in the
right panel. In principle, the strongest constraints would arise from the storage ring experiments involving protons and deuterons.
However, even in absence of these experiments, which are still on the drawing board, relevant constraints could be set by expected
improvements on $d_n$ and, interestingly, by future measurements of $d_\tau$ at Belle-II. 
A potential significant value for 
$g_T^c \neq 0$ again has a large impact. For example, taking $g_T^c=\frac{m_s}{m_c}g_T^s$ strengthens the constraints of hadronic EDMs (especially for the deuteron) and changes the combinations of $X_{\tau t}$ and $X_{\tau c}$ they are sensitive to.

\begin{figure}
\centering{
\includegraphics[scale = .56]{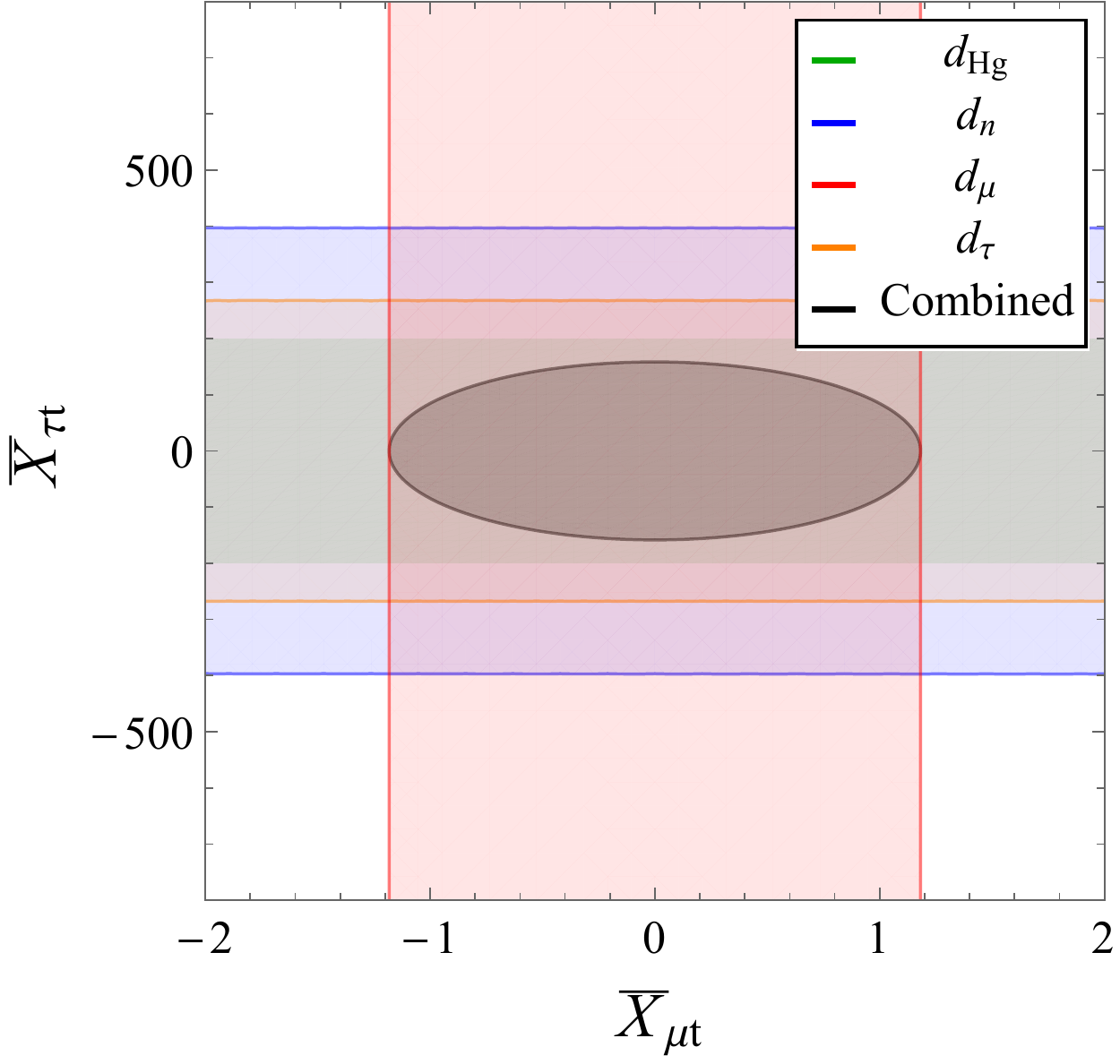}\hfill
\includegraphics[scale = .55]{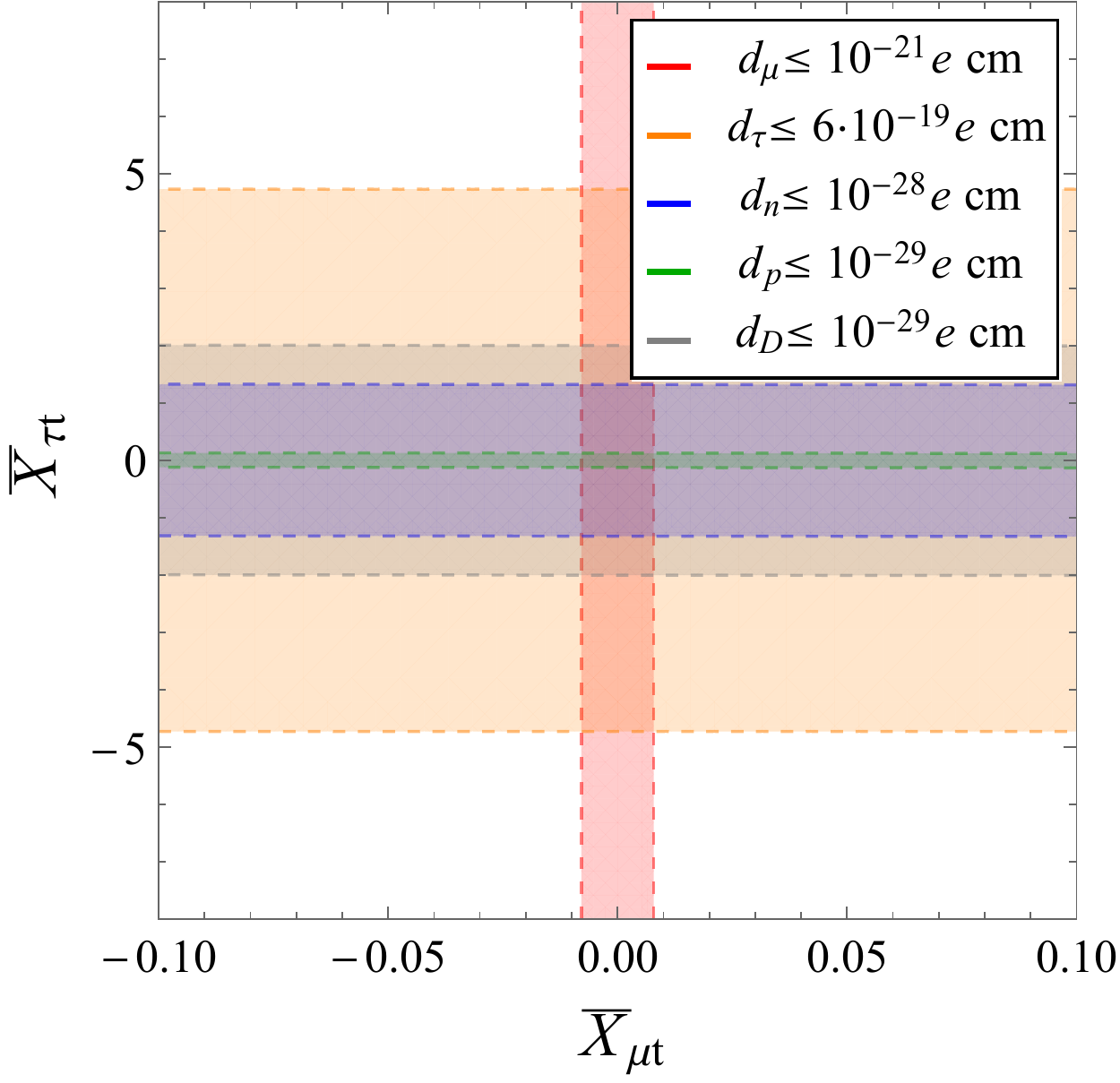}
}
\caption{Left panel: Constraints from various EDM experiments in the $\bar X_{\mu t}$-$\bar X_{\tau t}$ plane. The dark ellipse is the combined allowed region. Right panel:  constraints using the expected sensitivities of prospected muon, neutron, proton, deuteron, and $\tau$ EDM experiments.}
\label{LQ69}
\end{figure}

It is worthwhile to consider a scenario with only top couplings. In this case, the $\bar X_{e t}$ coupling is strongly constrained by
the ThO experiment, while the $\bar X_{\mu t}$ coupling is, at the moment and in the foreseeable future, only constrained by the limit
on $d_\mu$. On the other hand, $\bar X_{\tau t}$ is  constrained by $d_{\mathrm{Hg}},d_\tau$ and $d_n$, but only very weakly, and
in the future by $d_n$, $d_{\tau}$, $d_p$ and/or $d_D$,  as can be seen from Fig.\ \ref{LQ69}. As a result, all the top couplings can in principle
be constrained simultaneously.

\begin{figure}
\centering{
\includegraphics[scale = .55]{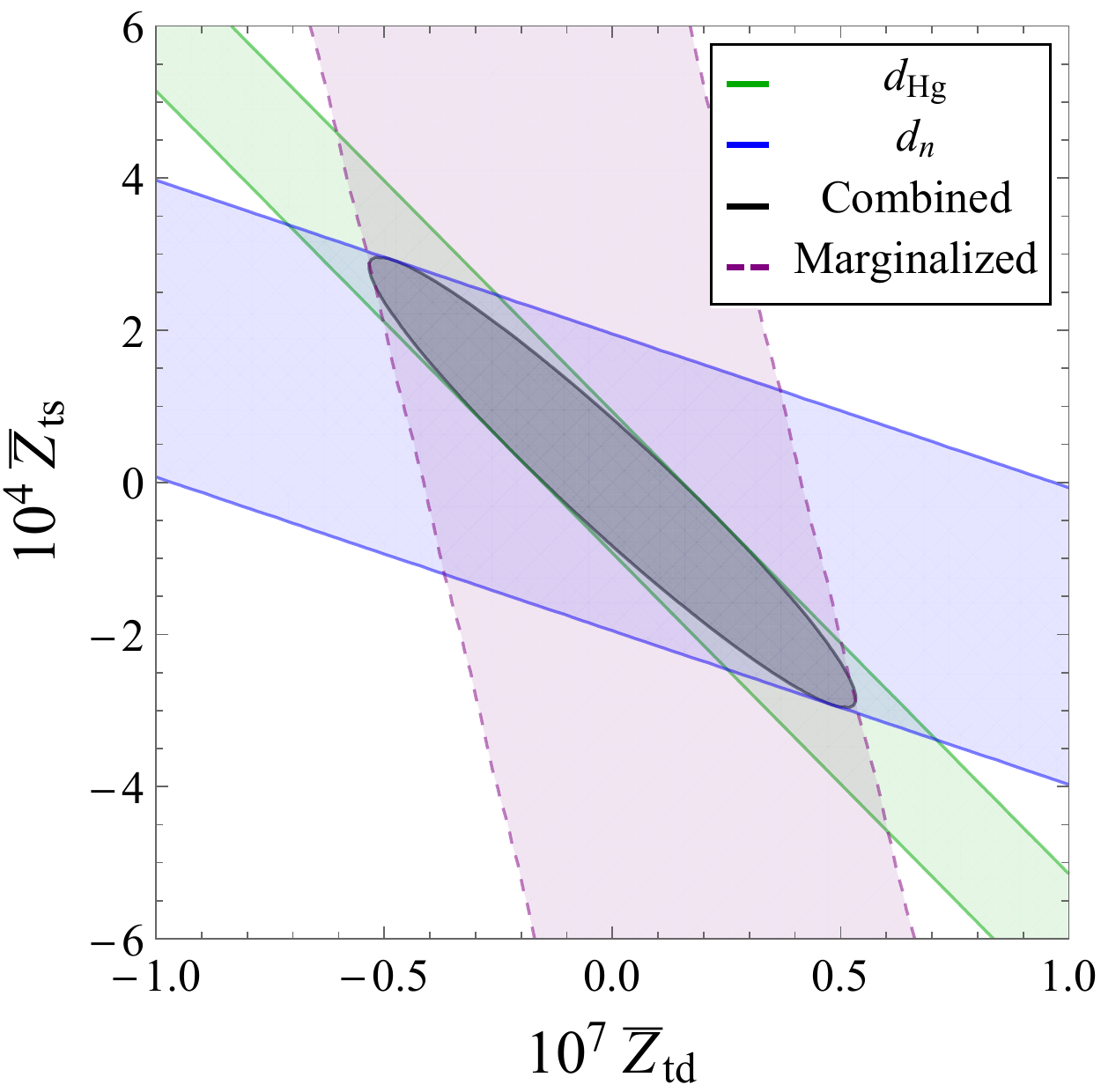}\hfill
\includegraphics[scale = .55]{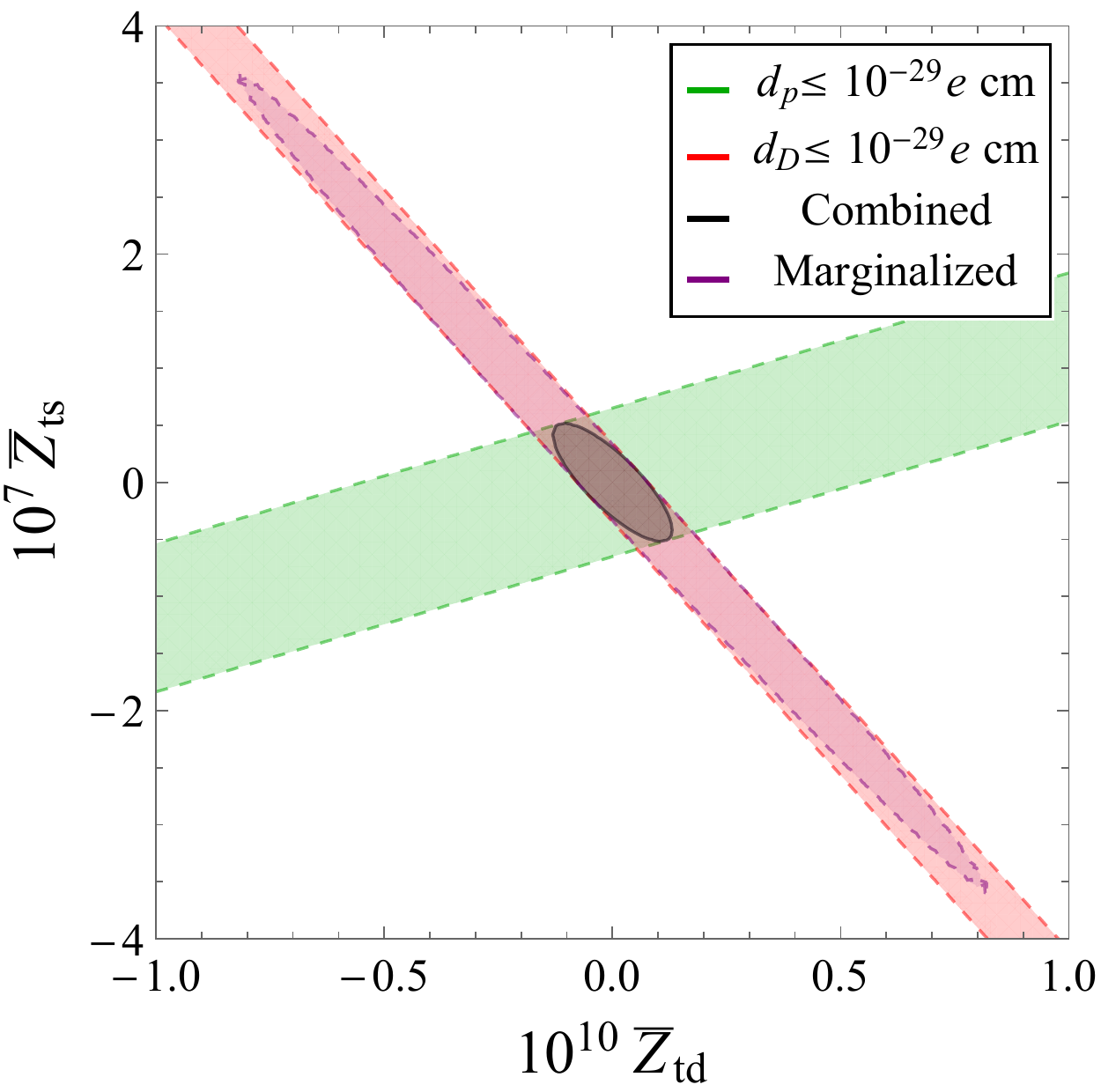}
}
\caption{Left panel: Constraints from various EDM experiments in the $\bar Z_{t d}$-$\bar Z_{t s}$ plane. The dark ellipse is the
combined allowed region, while the purple dashed band illustrates the allowed region when we also assume a nonzero $Z_{tb}$ coupling.
Right panel: constraints using expected sensitivities of future proton and deuteron EDM experiments. }
\label{LQZ79}
\end{figure}

A similar top scenario can be studied for the $\bar{Z}_{qq'}$ couplings. In Fig.~\ref{LQZ79} we show the  $\bar{Z}_{t
d}$-$\bar{Z}_{ts}$ plane. At present, only two EDMs are relevant, which are sufficient to constrain both couplings. The purple band,
however, shows that once we also turn on $Z_{tb}$ a free direction emerges, which would require additional measurements to constrain. 
Future  $d_p$ and $d_D$ experiments would both improve the limits and remove the free direction in the top sector as shown in the right
panel.

The above examples show that it is not possible to single out a single EDM experiment that is most important. Depending on the
couplings under consideration, essentially all EDM experiments play a role. While current limits on, for example, $d_\mu$ and $d_\tau$
are much weaker than limits on $d_n$, $d_\mathrm{Hg}$, and $d_e$ they are still important in constraining couplings involving muons and
taus.

\subsection{Lepton Flavor Universality violation in $B$ decays}
The experimental hints for lepton-flavor universality (LFU) violation in $B$ decays, most prominently reflected in the ratios
$R_{D,D^*}$ and $R_{K,K^*}$, have received much attention over the last few years. 
In trying to jointly understand both charged-current and neutral-current deviations, leptoquark models have emerged as uniquely
suited mediators. As an explicit example we discuss here a recently suggested model, involving two scalar leptoquarks, which has a
UV-completion based on $SU(5)$ Grand Unified Theory \cite{Becirevic:2018afm}. The model accommodates  the anomalies in  the $b\to c
\tau\bar \nu_\ell$ transitions ($R_{D^{(*)}}$) as well as those in  $b\to s \ell\bar\ell$ transitions ($R_{K^{(*)}}$) by introducing
$R_2$ and $S_3$ scalar leptoquarks, respectively.
While the interactions related to $S_3$ do not lead to significant effects in EDMs, the $R_2$ leptoquark generates a rich EDM
phenomenology as discussed here. Importantly, the model under consideration provides a direct link between effects
in $b\to c\tau\bar \nu_\ell$ transitions and EDMs.

We start by discussing how the $R_2$ couplings can accommodate the current anomalies in the $R_{D^{(*)}}$ ratios. 
These LFU ratios are defined as
\begin{equation}
R_{D^{(*)}} = \frac{\mathcal{B}(B\to D^{(*)} \tau \nu_\tau)}{\mathcal{B}(B\to D^{(*)} \ell \nu_\ell)} \ ,
\end{equation}
where $\ell = e, \mu$. Within the scenario of Ref.\ \cite{Becirevic:2018afm}, the $R_2$ leptoquark affects these ratios by modifying
the decays to $\tau$ leptons. The required $R_2$ couplings take the following form in our notation:
\bea\label{eq:ModelCouplings} x_{RL} = -\bma 0 &0&0\\0&y_L^{c\mu}&y_L^{c\tau}\\0&0&0\ema\,,\qquad x_{LR}^\dagger = V\bma 0 &0&0\\0&0&0\\0&0&y_R^{b\tau}\ema\, .
\eea
Below the electroweak scale,
corrections to $b\to c\ell \bar \nu_\ell$ transitions are induced by the following effective Lagrangian (in terms of the flavor eigenstates of the neutrinos),
\begin{eqnarray}\label{eq:efflag}
\mathcal{L}_{\rm eff} &=&- \frac{4 G_F}{\sqrt{2}} V_{cb} \left[ (\bar{c}_L \gamma^\mu b_L) (\bar{\tau}_L \gamma_\mu \nu_\tau)
+ g_{S_L}(\bar{c}_R b_L) (\bar{\tau}_R \nu_\tau) + g_T (\bar{c}_R \sigma^{\mu\nu} b_L) (\bar{\tau}_R \sigma_{\mu\nu} \nu_\tau) \right]  
\ ,
\end{eqnarray}
where the first term represents the SM contribution. The tensor and scalar terms are the charged-current pieces of the
$C_{lequ}^{(1,3)}$ operators in Eq.~\eqref{eq:Broken4Fermion}. The form of the $R_2$ couplings in  Eq.\ \eqref{eq:ModelCouplings},
together with the matching conditions in Eq.\ \eqref{eq:MatchFQsemil}, give rise to the following contributions:
\bea
g_{S_L}(m_{LQ}) =4g_{T}(m_{LQ}) =-\left(2\sqrt{2}G_F V_{cb}\right)^{-1}\left(C_{lequ}^{(1)\, \tau\tau q c}\right)^* V_{qb} = \frac{y_L^{c\tau}\left(y_R^{b\tau}\right)^*}{4\sqrt{2} G_F V_{cb} m_{LQ}\sq}\, .
\eea
The neutral-current part of the same operator includes
one of the combinations of couplings that contributes to EDMs, namely
\bea
{\rm Im }\, g_{S_L}(m_{LQ}) = -\frac{X_{\tau c}}{4\sqrt{2}G_F|V_{cb}|^2}\,.
\eea

The LFU ratios can be expressed in terms of the scalar and tensor couplings  in Eq.\ \eqref{eq:efflag} as follows
\cite{Feruglio:2018fxo}:
\begin{equation}\label{eq::RDDstar}
\frac{R_{D^{(*)}}}{R^{\text{SM}}_{D^{(*)}}} = 1 + a_{S_L}^{D^{(*)}}  |g_{S_L}(m_b)|^2 + a_T^{D^{(*)}} |g_T(m_b)|^2 + \tilde{a}_{S_L}^{D^{(*)}}  \text{Re}\; g_{S_L}(m_b) + \tilde{a}_T^{D^{(*)}}  \text{Re}\; g_{T}(m_b) \ ,
\end{equation}
where the coefficients $a^{D^{(*)}}_{S_L,T}$ contain phase space factors and form factor ratios and we use the numerical values derived
in Ref.~\cite{Feruglio:2018fxo}. In the above, all couplings are to be evaluated at $\mu=m_b$, for which one has,
\begin{equation}
1.64\, g_{S_L}(m_{LQ}) \simeq  g_{S_L}(m_b) \simeq 7.8 g_T(m_b) \ .
\end{equation}

The averages of the experimental measurements are \cite{Amhis:2016xyh,Lees:2012xj,Huschle:2015rga, Aaij:2017deq, 
Lees:2013uzd, Sato:2016svk, Aaij:2015yra, Aaij:2017uff,Hirose:2016wfn, Hirose:2017dxl} 
\begin{equation}\label{eq:RDmeas}
R_D^\text{exp} = 0.407\pm 0.046 \ , \;\;\; R_{D^*}^\text{exp} = 0.306 \pm 0.015\,,
\end{equation}
with a correlation of $20\%$, 
while the SM predictions are given by\footnote{Since we use Eq.~\eqref{eq::RDDstar} with coefficients from Ref.~\cite{Feruglio:2018fxo},
we use also their SM predictions.}
\cite{Feruglio:2018fxo}

\begin{equation}
R_D^\text{SM} = 0.293 \pm 0.007  \ , \;\;\; R_{D^*}^{\text{SM}} = 0.257\pm 0.003 \ .
\end{equation}
The prediction for $R_D$ is based on lattice-QCD results for the $B\to D$ form factors \cite{Lattice:2015rga,Na:2015kha}.
The form factors for $R_{D^*}$ are taken from Ref.~\cite{Bernlochner:2017jka}. 
The resulting predictions agree within uncertainties with
Refs.~\cite{Bigi:2016mdz,Bernlochner:2017jka,Bigi:2017jbd,Jaiswal:2017rve}.

\begin{figure}[t]
\centering{
\includegraphics[scale = .55]{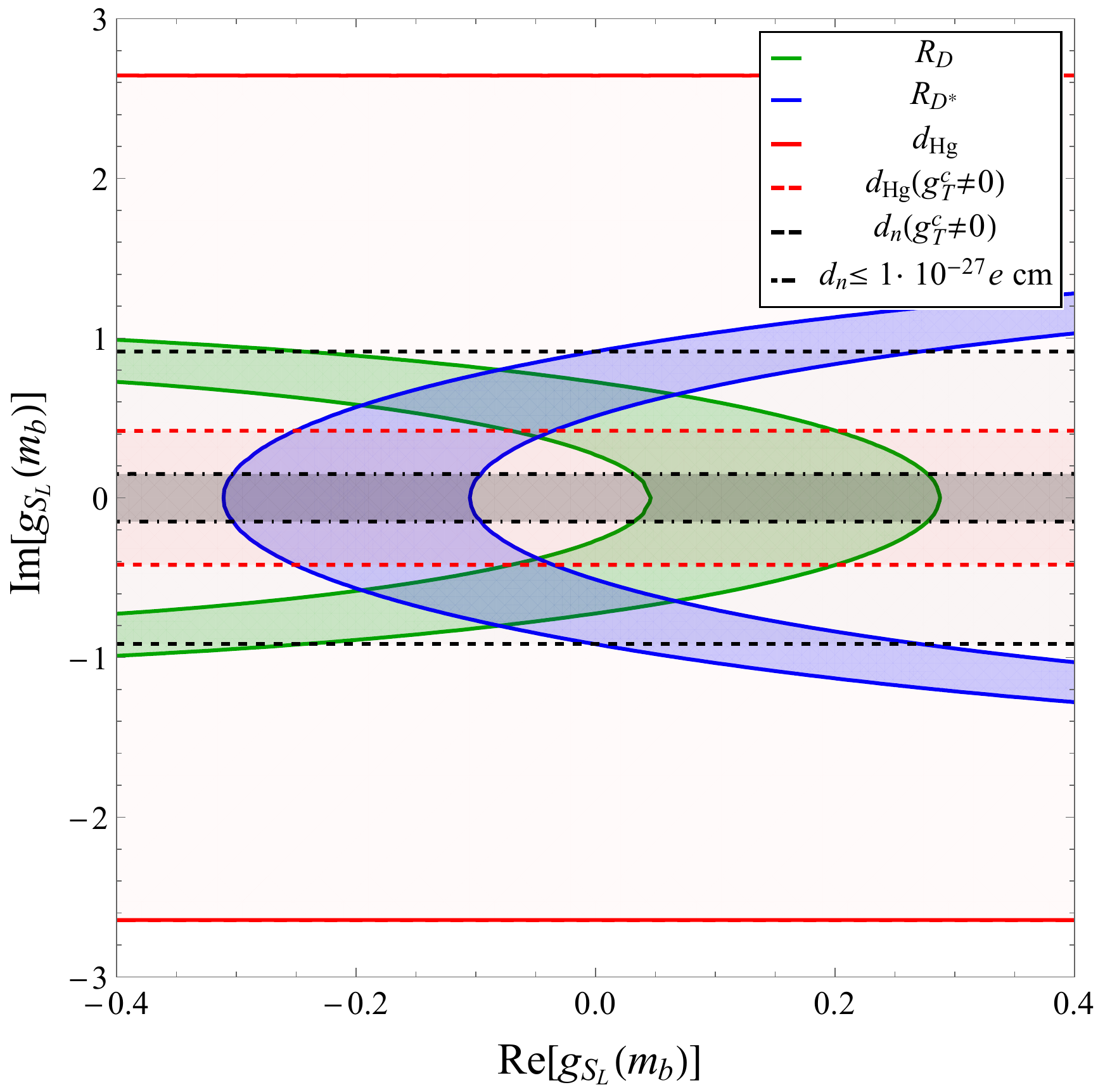}
}
\caption{Contours in the ${\rm Re}\; g_{S_L}(m_{b})$-${\rm Im}\; g_{S_L}(m_{b})$ plane. The $R_D$ and $R_D^{*}$ contours (at $90\%$ C.L.) are shown in  green and blue, respectively. The current constraints from from the  $\text{Hg}$ EDM is shown in red, while the dark-red band is a projection for a future neutron EDM measurement assuming an order of magnitude improvement. }
\label{fig:RDcons}
\end{figure}

Taking only the uncertainty on the experimental measurements of $R_{D^{(*)}}$ into account, we obtain the $90\%$ C.L.\ contours in
Fig.~\ref{fig:RDcons} in the $\mathrm{Re}\,g_{S_L}(m_b)- \mathrm{Im}\,g_{S_L}(m_b)$ plane. The SM point with $g_{S_L}(m_b)=0$ is
excluded at the few $\sigma$-level. 
The critical point is that a combined explanation of the $R_{D^{(*)}}$ anomalies requires
a nonzero $\mathrm{Im}\,g_{S_L}(m_b)$, in agreement with Ref.~\cite{Becirevic:2018afm}.
Since the imaginary parts of the couplings $g_{S_L}$  and $g_T$ are related to $X_{\tau c}$, they can be constrained by EDMs.\footnote{In principle additional contributions to EDMs can arise from diagrams that include both the $R_2$ and $S_3$ leptoquarks, which are not included in the analysis of Sect.\ \ref{Match}. However, such contributions can be shown to be suppressed by additional loop factors or small Yukawa couplings compared to those from $R_2$ alone.} 

There are two relevant contributions of $X_{\tau c}$ to EDMs: the first is a
sizable contribution to the three-gluon Weinberg operator via two-loop effects (see Fig.~\ref{Fig:RG} and Table~\ref{tab:RunningR2}).
The Weinberg operator in turn leads to nonzero nucleon EDMs and thus a nonzero $d_n$ and $d_{\mathrm{Hg}}$.
The other contribution is via the charm-quark EDM, again inducing nucleon EDMs and hence $d_n$ and $d_{\rm Hg}$. This contribution is
potentially much larger; in fact, for values of $g_T^c$ down to about $\sim 1/10$ of the present central value, this
constraint would rule out the values of $\mathrm{Im}\,g_{S_L}(m_b)$ required to explain the  $R_{D^{(*)}}$ measurements. 

The interpretation of  $d_n$ and $d_{\mathrm{Hg}}$ in terms of both $C_{\tilde G}$ and $d_c$ suffer presently from large
theoretical uncertainties; hence, it is too early to draw strong conclusions regarding the viability of this model. Since for the
central value of the Weinberg matrix element and $g_T^c = \frac{m_s}{m_c} g^s_T$ the constraint from $d_c$ is a factor $\sim 6$ stronger than that via
the Weinberg operator, we consider the latter a conservative constraint. In order to be able to easily adapt our results to future
determinations of $g_T^c$, we provide the following simplified formulae:
\bea
\left|{\rm Im}\,g_{S_L}(m_b)\right| &\leq&0.92 \left| \frac{2.2\cdot 10^{-4}}{g_{T,{\rm min}}^c(2{\rm
GeV})}\frac{d_n^{\rm limit}}{3.0\cdot 10^{-26}\, e\,{\rm cm}}
\right|\,,\\
\left|{\rm Im}\,g_{S_L}(m_b)\right| &\leq& 0.42\left| \frac{2.2\cdot 10^{-4}}{g_{T,{\rm min}}^c(2{\rm
GeV})}\frac{d_{\rm Hg}^{\rm limit}}{6.3\cdot 10^{-30}\, e\,{\rm cm}}
\right|\,,
\eea
where we assumed that both EDMs are dominated by the $d_c$ contribution.

From these observations,
one would expect next-generation $d_n$ or $d_{\rm Hg}$ experiments to see a signal if this particular model accounts for the $B$
anomalies. Further improvements of hadronic and nuclear theory would be very helpful to strengthen this conclusion.
We illustrate this situation in  Fig.~\ref{fig:RDcons}: there we show the constraints from $R_{D^{(*)}}$ in the complex $g_{S_L}$
plane together with the present bounds from $d_{\rm Hg}$ via the Weinberg operator and the charm EDM for $g_T^c=\frac{m_s}{m_c}g_T^s$. Additionally, for the neutron EDM, we show the current constraint using  $g_T^c=\frac{m_s}{m_c}g_T^s$ and the future limit assuming an improvement by a factor of 30 over the
current limit, $d_n < 1.0 \cdot 10^{-27}\, e$, but with $g_T^c=0$.

This example shows that EDMs can play a role in constraining leptoquark models that explain the $B$ anomalies, even if the latter
require flavor-changing interactions that do not directly lead to EDMs. In fact, in the above examples, $g_{S_L}$ and $g_T$ are induced
at tree level while EDMs are only induced at the two-loop level and suffer from an additional suppression of $V_{cb}^2 \simeq 1.7 \cdot
10^{-3}$. The fact that the EDM limits can still be relevant shows the power of EDM measurements in constraining new CP-violating
physics. It would be interesting to study other leptoquark solutions to $B$ anomalies and their EDM signature.

\section{Conclusions}\label{conclusion}

We have investigated how electric dipole moments of various systems are induced in models involving scalar leptoquarks. Depending on
their gauge representation, leptoquarks can possess both left- and right-handed interactions with fermions with a relative CP-violating
phase. We focused on two types of scalar leptoquarks, $R_2$ and $S_1$, where this is the case.
Other representations can also lead to EDMs but these require additional weak interactions and off-diagonal CKM elements. While EDMs
induced by $R_2$ and $S_1$ leptoquarks have been studied before \cite{Barr:1987sp, Geng:1990gr, Barr:1992cm, He:1992dc, Herczeg:2003ag,
Arnold:2013cva,Fuyuto:2018scm}, these studies focused on a subset of leptoquark interactions with light fermions. In this work, we
have generalized these results by including interactions to all quarks and leptons and show that this leads to a rich EDM phenomenology
and impressive constraints on CP-violating phases.

In order to avoid LHC constraints, we have assumed that potential leptoquarks are heavy with respect to the electroweak scale. We have
integrated out the leptoquarks and matched to CP-violating dimension-six operators of SM-EFT. These CP-violating operators consist of
electroweak and chromo-electric dipole operators, the Weinberg operator, and several four-fermion operators. The latter can be
lepton-quark, and, in case of $S_1$, quark-quark interactions involving all generations of quarks and leptons. We have evolved this set
of operators to the electroweak scale where we integrated out the heavy SM degrees of freedom and matched to CP-odd $SU(3)_c\times
U(1)_{\mathrm{em}}$-invariant operators (involving only 5 quark flavors). We subsequently evolved these interactions to the low-energy scales where EDM experiments
take place. All CP-odd operators that involve quarks or gluons require a matching to the hadronic level. We have performed this
matching based on chiral perturbation theory, using up-to-date hadronic matrix elements. Finally, we use the leptonic and hadronic
CP-odd interactions to evaluate EDMs of leptons, nuclei, atoms, and molecules using state-of-the art nuclear and atomic matrix
elements. We stress that several hadronic and nuclear matrix elements are still poorly known and include this uncertainty in our
analysis.

For leptoquark interactions involving electrons we find that all CP-phases are strongly constrained by EDM experiments involving
paramagnetic atoms and polar molecules. For couplings involving electrons and light quarks such EDMs are dominated by
semileptonic four-fermion interactions, while couplings among electrons and heavier quarks lead to large electron-EDM contributions.
For leptoquarks in the TeV-range, this leads to constraints on the imaginary parts of the relevant couplings at the $10^{-8,-9}$
level. Similar conclusions were recently reached in Ref.~\cite{Fuyuto:2018scm}.

For interactions involving muons and taus the situation is more complicated. 
Depending on the combination of leptoquark couplings and the statistical treatment, any of the 
experimental limits on the neutron, Hg, or lepton EDMs can lead to the strongest constraint.
In general, we conclude that leptoquark interactions involving second- and
third-generation leptons lead to a very distinct pattern of EDMs compared to interactions involving electrons. For the CP-odd
quark-quark interactions that appear in $S_1$ models the only relevant constraints arise from the neutron and Hg EDM
limits.
Here $d_{\rm Hg}$ gives the most stringent limits in the cases where large pion-nucleon interactions are induced (namely, for $Z_{us}$, $Z_{ub}$,
$Z_{td}$, and $Z_{cd}$). Instead the $d_n$ and $d_{\rm Hg}$ limits are comparable for the remaining couplings as they do not generate
enhanced pion-nucleon interactions.

All limits for a single CP-violating coupling are given in
Tables~\ref{tab:CentralLimitsR2}-\ref{tab:RfitLimitS1Quark}. These Tables also show how future EDM experiments involving different
nuclear and atomic systems would affect our conclusions. In Sect.~\ref{sec:interplay} we have investigated more realistic scenarios
involving more than one nonzero CP-odd interaction, which exemplify the complementarity of different (future) EDM experiments.
An important conclusion of our work is that all classes of EDM experiments (lepton, nucleon, nuclear, diamagnetic and
paramagnetic) play a role in limiting the various leptoquark interactions, motivating experimental and theoretical improvements on all
fronts.

To show potential applications of this work, we have applied our framework to a recent model of leptoquarks motivated by the anomalies
in $B$ flavor experiments \cite{Becirevic:2018afm}. Resolving these anomalies using leptoquarks generally requires interactions between
second- and third-generation quarks and leptons. If these interactions allow for CP-violating phases, as is the case in
Ref.~\cite{Becirevic:2018afm}, they can lead to EDMs. Using the results obtained here, we set limits on the complex couplings appearing
in this model. We find that this scenario remains consistent with current EDM experiments given the large theoretical
uncertainties specifically in $g_T^c$, but predicts a signal in the next generation of neutron EDM experiments. This example shows
that EDMs can play an important role in the study of leptoquark models.

\section*{Acknowledgments}
WD acknowledges support by the US DOE Office of Nuclear Physics and by the LDRD program at Los Alamos National Laboratory.
The work of MJ is supported by the DFG cluster of excellence “Origin and Structure of
the Universe”.
KKV~is supported by the Deutsche Forschungsgemeinschaft (DFG) within research unit FOR 1873 (QFET). We thank Andreas Crivellin for
useful discussions in the early phases of this work.

\bibliographystyle{h-physrev3} 
\bibliography{bibliography}

\end{document}